\let\accentvec\vec
\documentclass[iop]{emulateapj}

\let\vec\accentvec
\usepackage{amsmath}
\usepackage{amssymb}
\usepackage{graphicx}
\usepackage{hyperref}
\usepackage{tocbibind}

\begin{document}

\newcommand{\D}{\partial}
\newcommand{\DD}{\frac}

\title{Launching proton-dominated jets from accreting Kerr black holes: \protect\\
                     the case of M87}

\author{Brezinski\altaffilmark{1} F. \and Hujeirat, A.A.\altaffilmark{2}}

\altaffiltext{1}{ITP - Institut für Theoretische Physik, Philosophenweg 16, 69120 Heidelberg, Germany }
\altaffiltext{2}{ IWR - Interdisciplinary Center for Scientific Computing, University of Heidelberg, INF 368, 69120 Heidelberg, Germany}

\begin{abstract}
A general relativistic model for the formation and acceleration of low mass-loaded jets from systems containing accreting
black holes is presented. The model is based on previous numerical results and theoretical studies in the Newtonian regime,
but modified to include the effects of space-time curvature in the vicinity of the event horizon
of a spinning black hole.

It is argued that the boundary layer between the Keplerian accretion disk and the event horizon is best suited
for the formation and acceleration of the accretion-powered jets in active galactic nuclei and micro-quasars.

The model presented here is based on matching the solutions of three different regions:
i- a weakly magnetized Keplerian accretion disk in the outer part, where the transport of angular momentum is
mediated through the magentorotational instability, ii- a strongly magnetized, advection-dominated and
turbulent-free boundary layer (BL) between the outer cold accretion disk and the event horizon and
where the plasma rotates sub-Keplerian and iii-  a transition zone (TZ)
between the BL and the overlying corona, where the electrons and protons are thermally uncoupled, highly dissipative
and rotate super-Keplerian.

In the BL, the gravitation-driven dynamical collapse of the plasma increases the strength of the poloidal magnetic
 field  (PMF) significantly, subsequently suppressing the generation and dissipation of turbulence and
 turning off the primary source of heating.
 In this case, the BL  appears much fainter than standard disk models so as if the disk truncates at a certain
 radius. The action of the PMF in the BL is to initiate torsional Alf$\grave{v}$en waves that transport
angular momentum from the embedded plasma vertically into the TZ, where a significant fraction of the shear-generated toroidal
magnetic field reconnects, thereby heating the protons up to the virial-temperature. Also, the strong
PMF forces the electrons to cool rapidly, giving rise therefore to the formation of a gravitationally unbound two-temperature
proton-dominated outflow.

Our model predicts the known correlation between the Lorentz-factor and the spin parameter of the BH. It also shows that
the effective surface of the BL, through which the baryons flow into the TZ, shrinks with increasing the spin parameter,
implying therefore that low mass-loaded jets most likely originate from around Kerr black holes.

 When applying our model to the jet in the elliptical galaxy M87, we find a spin parameter $a \in[0.99, 0.998],$
 a transition radius $r_{tr}\approx 30 \text{~gravitational radii}$ and a fraction of $0.05\,-\, 0.1$ of the mass accretion rate
  goes into the TZ, where the plasma speeds up its outward-oriented motion to reach a Lorentz factor $\Gamma \in[2.5, 5.0]$
  at $r_{tr}.$ \\
\end{abstract}


\section{Introduction}

\begin{figure*}[ht]
\centering
		\includegraphics[bb=0 0 570 270,width=1.\textwidth]{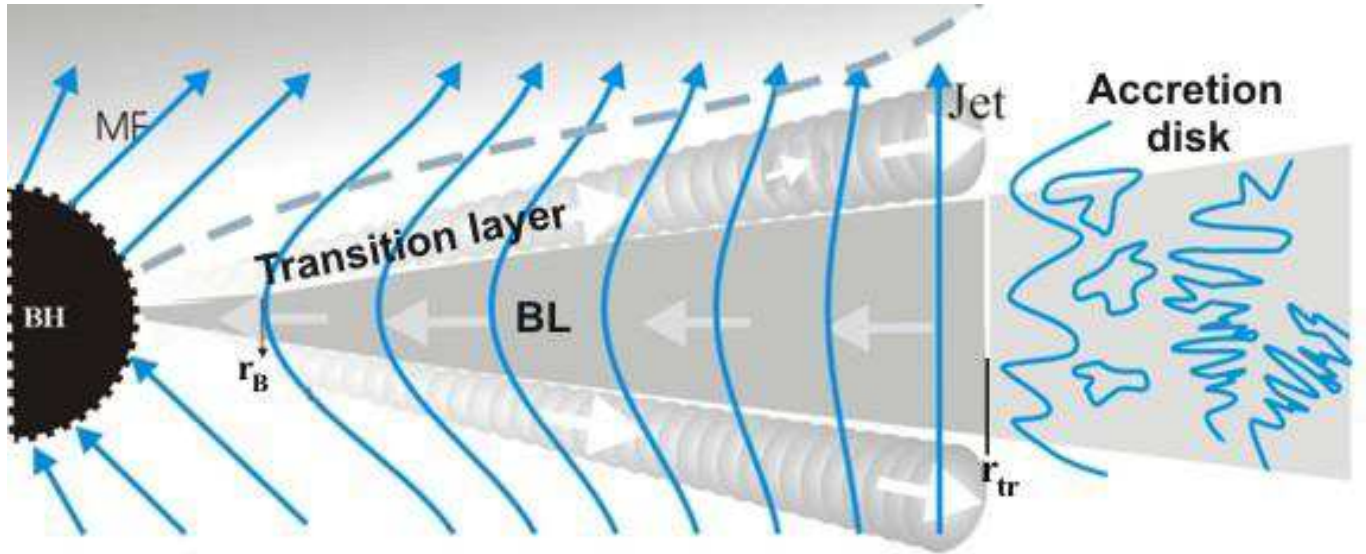}
	\caption{ A schematic description of the jet-disk-black hole interaction.
             At large radii the MF in the accretion disk is weak and disordered. The MF in the  innermost disk region, or equivalently in
             the boundary layer (BL) between the outer normal accretion disk and the black hole, are dragged inwards by the dynamical collapse of the plasma, where they become strong and of large scale topology (blue lines).
             Angular momentum transport in the BL is mediated by magnetic braking, thereby giving rise to the formation of a geometrically thin
              transition zone -TZ between the BL and the overlying tenuous
             corona.  The baryons in the TZ are dissipative and rotate with super-Keplerian velocity. The TZ in this model
             serves also as a runway for  accelerating the energetic baryons further by magnetic reconnection of the toroidal flux tubes.}
		\label{fig:schema}
\end{figure*}

Astrophysical jets have been observed to emanate from many astronomical systems such as around young stellar objects, in binary systems containing compact objects as well as in active galactic nuclei and quasars {{\citep[see][and the references therein]{camenzind,Livio2009}}}. Based on astronomical data and theoretical studies, the formation of jets in such systems is considered to be intimately connected to the accretion phenomena. These data reveal that the very collimated and fast propagating jets are found to emanate from systems containing black holes with some sort of correlation to their mass {{\citep[][]{FenderEtAl2007}}}.\\
In fact a lot of theoretical effort has been made to explain the jet-disk or jet-black hole connections, while the nature of the interaction between the black hole, jet and disk is  still not fully understood. A conclusive understanding of this interaction would require carrying out full three-dimensional, general relativistic, time-dependent, radiative magnetohydrodynamic calculations of dissipative plasmas, using a  multi-temperature and multi-component plasma description, which is beyond the numerical capability of the solvers available to date in
computational astrophysics.

{Alternatively, based on the combination of numerical, observational and theoretical results, we intend to construct a theoretical model that would describe the BH-jet-disk interaction properly. A similar approach has been presented by \citet{hujeirat2002,hujeirat2003,hujeirat2003-2,hujeirat2004}, though it applies for the Newtonian regime only.}
The present work is concerned with the extension of the above-mentioned Newtonian model  into the general relativistic regime. Such an extension is necessary, since the concerned region of interaction is located in the vicinity of the event horizon, {where the effects of the spin
and frame-dragging} are most prominent. Having performed these modifications, we may apply the model to study the formation and acceleration of relativistic jets both in micro-quasars and in active galaxies.\\
This paper is structured as follows: In Sect. \ref{qualmod} we establish the underlying assumptions and derive the governing equations in Sect. \ref{goveq}. Section \ref{solution} is dedicated to the derivation of the model and Sect. \ref{verification} to a verification of its self-consistency. In Sect. \ref{summary/outlook} we close with a summary and conclusions.

\section{Basics of jet-disk-black hole interaction}	\label{qualmod}

\begin{figure*}[ht]
\centering
	 \includegraphics*[width=12.5cm, height=6.5cm,clip] {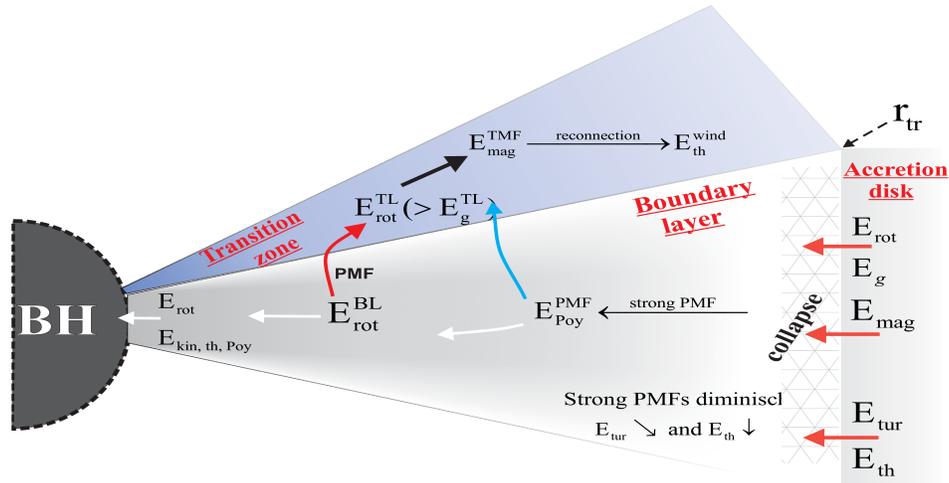}
\caption{{The rotational energy of the matter in Keplerian disks $\rm E_\mathrm{rot}$ is comparable to the gravitational energy,
$\rm E_\mathrm{g}.$ Part of $\rm E_\mathrm{rot}$
is converted via the MRI into magnetic and turbulent energies i.e. into $\rm E_\mathrm{mag}$ and $\rm E_\mathrm{tur},$ while the rest flows with the accreted matter across the transition radius $\rm r_\mathrm{tr}$ into the BL. $\rm r_\mathrm{tr}$ is defined to be the location, where magnetic and thermal energies become comparable i.e.,
$\rm E_\mathrm{mag}  \approx \rm E_\mathrm{th}.$
 The plasma in the BL undergoes a dynamical collapse, through which the magnetic energy of the poloidal magnetic field
$\rm E^\mathrm{PMF}_\mathrm{Poy}$ increases inwards as $\rm r^{-4}.$ This in turn splits $\rm E_\mathrm{rot}$ into a major part that goes into the transition zone, $\rm E^\mathrm{TZ}_\mathrm{rot},$ whereas
the rest is being accreted by the black hole in the form of kinetic energy $\rm E_\mathrm{kin}$. Moreover, the strong PMF diminishes  $\rm E_\mathrm{tur}$ and therefore $\rm E_\mathrm{th}.$ In the transition zone,
a part of  $\rm E^\mathrm{TZ}_\mathrm{rot}$ is converted into toroidal magnetic energy $\rm E^\mathrm{TMF}_\mathrm{mag}$ which in turn is
partially converted into thermal energy $\rm E^\mathrm{wind}_\mathrm{th}$ through magnetic reconnection to thermal-support the
outflowing wind. }}
\end{figure*}

{Consider the innermost part of an accretion disk around a central black hole, where the black hole's gravitational force  dominates the dynamics of the flow. At large distance, the flow is well described by the standard $\alpha$-disk model \citep[see][]{ssd}. Small scale magnetic fields (henceforth MF) were found to be amplified by the magneto-rotational instability -MRI \citep{mri1}, which is capable of
 turning initially laminar into turbulent flows. The generated turbulent eddies have the effect of viscosity, which converts
 a considerable part of the kinetic energy into black body radiation that can be observed in the UV and soft X-ray bands. Still, turbulent eddies and magnetic reconnection would maintain the magnetic energy to remain subequipartitioned with thermal energy, so that the ratio of the
 magnetic to gas pressure:  $\rm \beta = P_{mag}/P_{gas} $ is less than unity almost everywhere in the disk.
 {Following \cite{Lovelace2008}, weak magnetic fields do not hinder accretion, even when they are of large scale structure and
 threading both the corona and the disk.}

  However, there is no reason to expect $\beta < 1$ in the innermost part of the disk, where the effect of the spin of the central accreting
  black hole becomes significant \citep[see][and the references therein]{Punsly2009}. MRI, Parker instability, reconnection in combination with the significant speed up
  of inflow may lead to the establishment of a large-scale poloidal MFs, whose corresponding energy could be
   comparable
  or even larger than the thermal energy of the embedded plasma \citep{hujeirat2003}.}
  { In this case, strong MFs in the boundary layer (BL)  would suppress the generation of turbulence and therefore switch off the
   the source of local heating. Thus the BL would appear much fainter than standard disk models, or
   equivalently, the disk appears to truncate at a certain radius close to the central object \citep{Belloni2000, Hujeirat2000}.\\
  Moreover, regions governed by extremely strong magnetic fields are generally matter-free.
  {\citet{McKinney2004} and \citet{Fragile2008} argued that such Poynting flux dominated regions may form
  close to the BH and act as a launching mechanism in the sense of Blandford \& Znajek extraction process \citep{BZ77}.}
  Although ideal MHD solvers are incapable of modeling matter-free and magnetic dominated funnels, their simulations
   show that these form at rather {high latitudes and therefore most likely initially conditioned with no dynamical}
   coupling  to the accretion flow in the equatorial region.}
   {In our case however, the collapse-generated strong poloidal magnetic fields in the BL remains confined to the transition zone and does
   not diffuse throughout the corona, so that they effectively thread a small portion of the corona only.
   As it will be shown in Sec. (4), the time scale for the PMF to diffuse throughout the corona is much longer than  the time scale characterizing
   the dynamics of the plasma in the TZ.}
   {Noting that in the the absence of heating from below, rotational-unsupported corona around BHs are thermally unstable \citep{hujeirat2002},
   and that thermal conduction across the PMF lines is much weaker than along the field lines, we conclude that BH-corona most
   likely are relatively cold voids and therefore are thermally and dynamically unimportant for both the jet and the disk.  }\\ \\

The question to be addressed here is: what would be the most appropriate configuration of plasma flows with $\beta \geq 1$
   in the vicinity of rotating black holes  and whether such configurations could lead to the formation of the
     highly energetic jets observed to emanate from systems containing accreting black holes?\\

Let $\rm{r_{tr}}$ be the radius which separate two regions: exterior to $\rm{r_{tr}}$ the flow is said to obey the standard $\alpha$-disk
description. Interior to $\rm{r_{tr}},$ MFs are predominantly poloidal and of large-scale topology, where ideal MHD approximation holds.
In this case the magnetic field lines will be dragged inwards with the plasma particles under the action of the gravitational force of the
central BH. As the inflowing plasma rotates, angular momentum would be extracted and transported vertically on the dynamical
time scale. This time scale is comparable to $\rm t_A = H_d / v^\theta_A,$ where $H_d$ is the disk half-thickness and $\rm v^\theta_A$ is the Alfv\'en velocity along poloidal field lines.  A larger $\rm{r_{tr}}$ implies that more rotational energy would be extracted from the disk.
A too large $\rm{r_{tr}}$ however, would force the inner disk to start collapsing at lager radii, hence the inwards-drifted MFs may become sufficiently strong to terminate accretion completely as in the case of "magnetically arrested accretion", that has been reported by
\cite{Igumenshchev2008}.

\begin{figure}[ht]
\centering
		\includegraphics[angle=0,width=.5\textwidth]{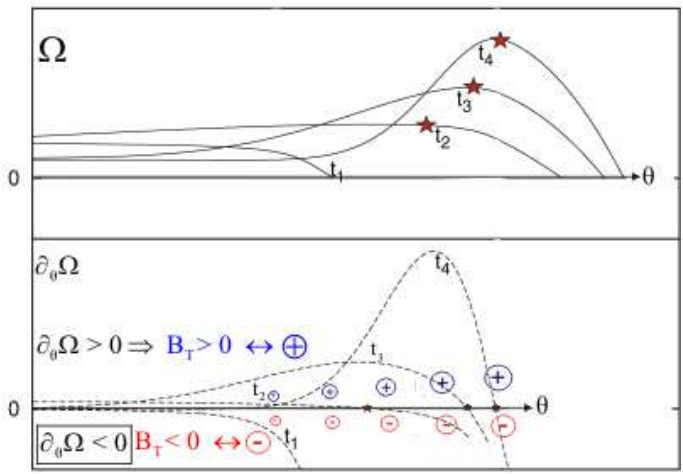}
\caption{A schematic description of the time-evolution of the angular velocity $\Omega$ and its derivative $\partial_\theta\Omega$ along the polar coordinate $\theta$ across the bounday layer for the time sequence $\rm t_1 < t_2 < t_3 < t_4$. The angular momentum of the plasma in
the BL is transported predominantly in the vertical direction by magnetic braking, thereby forming a thin transition zone between the BL and the overlying
tenuous corona, where the angular velocity peaks out. On the other hand,  since $\partial_\theta\Omega$ becomes negative as particles move
 from the disk to higher latitudes, and since $\rm B_\varphi \propto \partial_\theta\Omega$, we expect the TZ to accommodate a large
 number of flux tubes of opposite polarity, therefore giving rise to an enhanced magnetic reconnection.
 }
		\label{fig:vertical_omega2}
\end{figure}

 Indeed, our model predict that $\rm{r_{tr}} \leq 20 \, r_S$, where $\rm r _S$ is the Schwarzschild radius \citep{hujeirat2003,hujeirat2004}. This is much smaller than the typical dimension of the surrounding accretion disc, hence the importance of studying
 the innermost boundary layer (BL) between the disk and the central BH.\\
Supplied with angular momentum from the BL, the matter may start to rotate faster as it moves vertically in the manner
  depicted in Fig. (\ref{fig:vertical_omega2}). On the other hand, inspection of the evolution equation of the toroidal magnetic flux
   implies that the generation of - $B_\varphi$ is proportional to $\partial_\theta \Omega.$ As this gradient change signs at the
   interface betwwen the disk and the corona,  toroidal magnetic flux tubes of flux opposite orientation must be generated.
   Part of these tubes will subsequently intersect and reconnect, hindering thereby the transport of angular momentum to higher latitudes.
   The trapped rotational energy in the transition zone (TZ) between the disk and the overlying corona forces the plasma to rotate super-Keplerian and therefore become gravitationally unbound, hence starts to accelerate outwards.

    The characterizing features of our model compared to other theoretical and numerical models
     are two folds: 1- Formation of a geometrically thin turbulent-free region between the standard disk and the event horizon, where the
     plasma components are thermally coupled but rotate sub-Keplerian, and 2- Formation of two geometrically thin runaway zones
     that sandwich the above-mentioned turbulent-free region. The plasmas in these zones are proton-dominated, dissipative, two-temperature
     and super-Keplerian rotating. \\
     We note that simulations that rely on solving the ideal MHD equations alone are not not capable of capturing these features
     properly  \citep[e.g.][]{McKinney2004}.  On the other hand, sophisticated numerical calculations that relies
       on using the highly robust 3D axi-symmetic implicit Newtonian radiative MHD solver - IRMHD3 has been carried out
       by \citep{hujeirat2003}. These calculations have explicitly confirmed the formation of the above-mentioned features and
       predicted that a fraction of roughly $\rm \dot M_w/\dot M_d \approx 1/20$ of the inflowing matter goes into
       gravitationally unbound wind.

 In the present model, the matter in the TZ is provided by the inner disk by means of slow vertical drifts, $\rm v^\theta$, mediated by
  the thermal pressure and MFs. $\rm v^\theta$ is much slower than the other two velocity components. Due to the high reconnection rate and inefficient cooling, the protons in the TZ may easily be heated to reach the virial temperature. The super-Keplerian rotating plasma
  in the TZ is centrifugally accelerated outwards. A fraction of the toroidal MF in the TZ is advected with the wind. At a larger distance
   the geometrically diverging outflow ceases to be diffusive, where Lorentz-forces start to redirects the motion of the outflowing
   proton-dominated plasma to form a collimated jet.\\
Numerical calculations of \citet{hujeirat2002} have shown that a fraction of roughly $\rm \dot M_w/\dot M_d \approx 1/20$ of the inflowing matter goes into the form of a wind.

These properties are the foundation of the present model. The subject of the next two Sects. will be to deliver an analytic description within the context of general relativity.

\section{The governing equations}	\label{goveq}

In this Sect. we quote the equations governing the accretion flow. The mass of the central black hole is assumed to be much larger than the total mass in the surrounding accretion disk and thus self-gravitation of the flow is negligible. As a consequence we can describe spacetime by means of the Kerr metric. The equations of motion of the plasma are derived from the condition of conservation of energy, i.e. of the vanishing divergence of the stress-energy tensor, $\rm \nabla \cdot T = 0$, which consists of four independent equations. The three spatial components yield the momentum equations while the energy equation is given by the projection on the timelike four-velocity of the plasma, $\rm \dot x \cdot \left( \nabla \cdot T \right) = 0$. No viscous contributions to the stress-energy tensor are considered, since, in the present model, these are suppressed by strong MFs in the inner region of the accretion disk. In the absence of viscosity, there are three forms of energy that contribute to the stress-energy tensor which can then be written as
\begin{equation}
	\rm T = T_P + T_H + T_{EM},	\label{ergmom}
\end{equation}
where $T_P$, $T_{EM}$ and $T_H$ represent the stress-energy tensor of a perfect fluid as well as contributions due to electromagnetic fields (henceforth EMF) and thermal processes, respectively. They read as follows:
\begin{align}
	\rm {T_P}^\mu{}_\nu & = \rho \frac{\mathcal H}{c^2} \dot x^\mu \dot x_\nu + \delta^\mu{}_\nu P,	\label{perfect-stress-energy}	\\
	\rm {T_H}^\mu{}_\nu & = q^\mu \frac{\dot x_\nu}{c^2} + \frac{\dot x^\mu}{c^2} q_\nu,	\\
	\rm {T_{EM}}^\mu{}_\nu & = \frac 1 {4\pi} \left( F^{\mu\rho}F_{\nu\rho} - \frac 1 4 \delta^\mu{}_\nu F^{\alpha\beta}F_{\alpha\beta} \right).	 \label{em-stress-energy}
\end{align}
$\rm c$, $\rho$ and $\rm \dot x = dx/d\tau$ are the speed of light, the rest-mass density and the four-velocity of the plasma with proper time $\tau$, respectively. $\rm \mathcal H = c^2 + \mathcal E + P/\rho$ corresponds to the specific enthalpy, $\mathcal E$ to the internal energy per mass and $\rm P$ comprises gas, radiation and turbulent pressure but not magnetic pressure. $q$ is the heat flux vector which describes energy fluxes caused by various heating and cooling processes. It is purely spatial i.e. perpendicular to the fluid worldlines: $\rm \dot x \cdot q = 0$. The most relevant processes in accretion flows are cooling through bremsstrahlung, comptonization and synchrotron radiation as well as heating by viscous dissipation and magnetic diffusivity. Other processes that cause heat flux are Coulomb coupling between the electrons and protons, adiabatic compression and heat conduction \citep{hujeirat2004}.\\
The electromagnetic stress-energy tensor accounts for the energy content of the electromagnetic field (EMF), stresses exerted on the fluid by Lorentz forces and ohmic heating caused by electric currents running through a resistive plasma. This part is expressed in terms of the electromagnetic field strength tensor $F$. In order to take EMFs into account the system of equations has to be augmented by Maxwell's equations. In terms of $F$ they read
\begin{align}
	\nabla\cdot F & = - \frac{4\pi}{c} j,	\label{maxi}	\\
	\nabla\cdot *F & = 0,	\label{maxh}
\end{align}
where $(*F)_{\mu\nu} = \frac{1}{2} \varepsilon_{\mu\nu\rho\sigma} F^{\rho\sigma}$ is dual field strength tensor, $j$ the four-current density and $\varepsilon$ the totally antisymmetric Levi-Civita symbol. Before we proceed to the divergence of the stress-energy tensor, we introduce Ohm's law in order to derive the general relativistic induction equations. Unfortunately Ohm's law can be very complicated in general relativity \citep[for a detailed treatment see][]{grohm}. In order to obtain an analytic solution to the GRMHD equations, we have to settle for its most simple and straightforward approximation,
\begin{equation}
	\frac{4\pi}{c} \eta_M j^\nu = - \dot x_\mu F^{\mu\nu} \label{ohm},
\end{equation}
where $\eta_M = c^2/(4\pi \sigma)$ is the magnetic diffusivity and $\sigma$ the electric conductivity. This is the natural generalization of the Newtonian formula $\frac{4\pi}{c} \eta_M \vec j = c \vec E + \vec v \times \vec B$.\\
Using Eq. (\ref{maxh}) the resistive induction equations may be written in the form of an evolution equation for $F$,
\begin{equation}
	\frac{dF_{\mu\nu}}{d\tau} = \dot x^\rho \nabla_\rho F_{\mu\nu} = 2 F_{\rho[\mu} \nabla_{\nu]} \dot x^\rho - 2 \nabla_{[\mu} \left( \dot x^\rho F_{\nu]\rho} \right),	\label{grmhd'}
\end{equation}
where $[\,]$ indicates antisymmetrisation. Eq. (\ref{grmhd'}) contains three equations determining the evolution of the electric and MF components, respectively. Yet it will be much simpler to use Eq. (\ref{ohm}) to recover the electric fields.\\
Introducing the continuity equation
\begin{equation}
	\nabla_\mu \left( \rho \dot x^\mu \right) = 0	\label{cont}
\end{equation}
and making use of eqs. (\ref{maxi}), (\ref{maxh}) and (\ref{ohm}) one can derive the energy equation from $\dot x \cdot \left( \nabla \cdot T \right) = 0$,
\begin{equation}
	\rho \nabla_{\dot x} \mathcal E	=	- P \Theta + \frac{4\pi}{c^2} \eta_M j^2 + \sum_i \left( \pm \Lambda_i \right)	\label{grergeq},
\end{equation}
where $\Theta = \nabla \cdot \dot x$ is the four-dimensional volume expansion. The first term on the right-hand side of Eq. (\ref{grergeq}) accounts for heat generation by compression and the second describes ohmic heating by electric currents running through the fluid. The $\Lambda_i$ are heating- and cooling functions that represent all contributions from the heat flux vector $q$, namely
\begin{equation}
\begin{array}{lcl}
	\Lambda_{con}	& \hat = &	\mbox{heat conduction}	\\
	\Lambda_{pe}	& \hat = &	\mbox{electron-proton Coulomb coupling}	\\
	\Lambda_{syn}	& \hat = &	\mbox{cooling by synchrotron emission}	\\
	\Lambda_B			& \hat = &	\mbox{Bremsstrahlung-cooling}	\\
	\Lambda_C			& \hat = &	\mbox{cooling by componization}
\end{array}
\end{equation}
The term $\frac{4\pi}{c^2} \eta_M j^2$ in Eq. (\ref{grergeq}) will also be expressed in terms of a heating function, $\Lambda_{Ohm}$, in regions where we the ideal MHD approximation is invalid.\\
The three spacelike components of $\nabla\cdot T$ yield the momentum equations. In the vicinity of the central object gravity is the most dominant force. We will therefore neglect thermal contributions to the equations of motion, i.e. we set $\mathcal H = c^2$ and $q = 0$ in the momentum equations \citep[similar to the condition of negligible specific heat in][]{relsd,relsd2}. Using Eq. (\ref{cont}) we obtain
\begin{equation}
	\rho \nabla_{\dot x} \dot x_\nu = - \nabla_\nu P + \frac{1}{c} F_{\nu\rho} j^\rho	\label{greuleq}.
\end{equation}
Now we have derived the complete set of the GRMHD equations. In order to perform any specific calculations, however, we will have to rewrite them into a more practical form. Let $g$ be the metric, then the GRMHD equations read:\\

the continuity equation:	\label{mhdeqs}

\begin{equation}
	\frac 1 {\sqrt{\left| \det g \right|}} \partial_\mu \left( \sqrt{\left| \det g \right|} \, \rho \dot x^\mu \right) = 0,
\end{equation}

the momentum equations:

\begin{equation}
	\rho \dot x^\mu \partial_\mu \dot x_\alpha = \frac 1 2 \rho \dot x^\mu \dot x^\nu \partial_\alpha g_{\mu\nu} - \partial_\alpha P + \frac{1}{c} F_{\alpha\rho} j^\rho,
\end{equation}

the energy equation:

\begin{equation}
	\rho \dot x^\mu \partial_\mu \mathcal E = - P \Theta + \sum_i \left( \pm \Lambda_i \right),
\end{equation}

the induction equations:

\begin{multline}
	\dot x^\rho \partial_\rho F_{\mu\nu} = F_{\rho\mu} \partial_\nu \dot x^\rho - F_{\rho\nu} \partial_\mu \dot x^\rho -	\\
	- \frac{4\pi}{c} \left( \partial_\mu \left( \eta_M j_\nu \right) - \partial_\nu \left( \eta_M j_\mu \right) \right),
\end{multline}

the electric field equations (Ohm's law):

\begin{equation}
	\dot x_t F^{t\nu} = - \dot x_i F^{i\nu} - \frac{4\pi}{c} \eta_M j^\nu.	\label{ohmslaw}
\end{equation}
Now the GRMHD equations have been rewritten into a form that can be readily applied to any specific model.\\
In the following it is assumed that the central object is a rotating black hole so spacetime can be described by the Kerr-metric. This, however, is not a requirement of the model, which is independent of the central object in its basic assumptions. Further, we assume that the accretion disk lies in the equatorial plane and the system is reflection-symmetric. The coordinate system used is identical to the pseudo-spherical Boyer-Lyndquist coordinate system except that the polar coordinate $\theta$ is redefined by $\theta \rightarrow \pi/2 - \theta$. In the Newtonian model of \citet{hujeirat2004} the domain of interest was geometrically thin, thus we stay close to the equatorial plane: $\theta \approx 0$. Under these circumstances the Kerr-metric can be approximated by
\begin{equation}
	g = - \alpha^2 c^2 dt^2 + \varpi^2 \left( d\varphi - \omega dt \right)^2 + \frac{r^2} \Delta dr^2 + r^2 d\theta^2,	\label{metric}
\end{equation}
where the metric functions in the equatorial plane, correct up to order $\theta^2$, read
\begin{align}
	\alpha &= \frac{\sqrt\Delta}{\varpi},	&	\varpi &= \sqrt{r^2 + \rm{r_g}^2 a^2 + \frac{2 \rm{r_g}^3 a^2}{r}},	\nonumber	\\
	\omega &= \frac{2 \rm{r_g}^2 c a}{r \varpi^2},	&	\Delta &= r^2 - 2 \rm{r_g} r + \rm{r_g}^2 a^2,
\end{align}
where $\rm{r_g} = GM/c^2$ is the gravitational radius, $M$ is the mass of the central object and $a \in [-1,1]$ is the non-dimensional Kerr-parameter.\\
In the following we use the tetrad system of the zero angular momentum observer \citep[ZAMO, see e.g.][]{camenzind}. The components of the EMF as well as the current density, velocity and Lorentz factor are understood as measured by ZAMO. For the sake of clarity, however, the explicit expressions of these quantities, the sets of basis vectors and one-forms are given in the Appendix. 

\section{Constructing the combined solution}	\label{solution}

In this Sect. we describe the construction of the combined solution for both the BL and the TZ in the vicinity of the central object. {The solution is given based on the time independent and axisymmetric ($\partial_t = \partial_\varphi = 0$) equations. $v^\theta$ is taken to be much smaller than $v^r,v^\varphi$ and will be neglected where this is appropriate.}\\
Let $H_d$ be the disk half thickness in the BL and $H_w$ the thickness of the TZ. Both the BL as well as the TZ are assumed to be geometrically thin, i.e. $H_{d/w} \ll r$. The relative half thickness $H_d/r$ of the BL is further assumed to be constant, while the thickness of the TZ is allowed to vary with radius,
\begin{align}
	H_d/r &= \sin\theta_d \approx \theta_d,	\\
	H_w/r &= \sin\theta_w(r) - \sin\theta_d \approx \theta_w(r) - \theta_d,	\label{TLthickness}
\end{align}
where $\theta_d$ is a constant while $\theta_w$ is an unspecified function of $\rm r $. Defining the surface densities $\Sigma_d$ and $\Sigma_w$,
\begin{align}
	\Sigma_d &= \int\limits_{- \theta_d}^{\theta_d} d\theta \, r \rho \approx 2 H_d \rho,	&	\Sigma_w &= \int\limits_{\theta_d}^{\theta_w} d\theta \, r \rho \approx H_w \rho,
\end{align}
as well as the total mass $M$ inside a given three-dimensional, spacelike hypersurface $V$, averaged over the infinitesimal time slice $\delta t$,
\begin{equation}
	M := \frac{1}{\delta t} \int\limits_{t}^{t+\delta t} dt \int\limits_{V} dr d\theta d\varphi \, r^2 \, \rho \frac{\gamma}{\alpha},	 \label{relmass}
\end{equation}
one can derive {from the continuity equation,}
\begin{equation}
		\frac 1 {r^2} \partial_r \left( r \rho \sqrt\Delta \gamma v^r \right) + \frac 1 r \partial_\theta \left( \rho \gamma v^\theta \right) = 0,	 \label{continuityequation}
\end{equation}
{the well known expressions} for the accretion rate, $\dot M_d$, in the disk and the outflow rate, $\dot M_w$ in the TZ,
\begin{align}
	\dot M_d &= - 2\pi \sqrt\Delta \Sigma_d \gamma_d v^r_d,	\label{mdotdisk}	\\
	\dot M_w &= 2\pi \sqrt\Delta \Sigma_w \gamma_w v^r_w,	\label{mdotwind}
\end{align}
where $\gamma$ and $v^r$ have been approximated by their vertical averages in the integration process. The TZ is fed with matter from the disk in the BL by means of a small vertical drift $v^\theta$. Correspondingly $\dot M_d$ and $\dot M_w$ are neither independent of each other nor constant with radius but obey the relation
\begin{equation}
	\partial_r \dot M_d(r) = 2 \partial_r \dot M_w(r),	\label{drmdot}
\end{equation}
where the factor of two accounts for the two surfaces of the disk. $\dot M_d$ and $\dot M_w$ further obey the boundary conditions
\begin{align}
	M_d(\rm{r_{tr}}) &= \dot M,	&	M_d(r_{B}) &= \dot M_{d,B},	\\
	2 M_w(\rm{r_{tr}}) &= \dot M \dot{\mathcal M_{tr}},	&	M_w(r_{B}) &= 0,	\label{bcs}
\end{align}
where $\dot M$ corresponds to the total accretion rate at $\rm{r_{tr}}$ and the relative outflow rate in the TZ is defined by
\begin{equation}
	\dot{\mathcal M}(r) = 2\dot M_w(r)/\dot M.	\label{relout}
\end{equation}
For the moment the derivation of $\dot{\mathcal M}$ and hence the profiles of $\dot M_d$ and $\dot M_w$ is postponed since it requires prior knowledge of several solutions which we have to derive first.\\

Now we turn to the basic assumptions that \citeauthor{hujeirat2004} made for his Newtonian model. The first assumption is that the ideal MHD approximation holds inside the BL of the accretion disk. {In the non-relativistic limit the induction equations read
\begin{equation}
	0 = \vec{\nabla} \times \left( \vec{v} \times \vec{B} - \frac{4\pi}{c} \eta_M \vec{j} \right).
\end{equation}
Correspondingly, the poloidal components are given by
\begin{align}
	0 &= \partial_\theta \left( r v^r B^\theta - \frac{4\pi}{c} \eta_M j_\varphi \right),	\label{nrradb}	\\
	0 &= - \partial_r \left( \varpi \gamma v^r B^\theta - \frac{4\pi}{c} \eta_M j_\varphi \right).	\label{nrvertb}
\end{align}
The relativistic generalization of eqs. (\ref{nrradb}) and (\ref{nrvertb}) is simply given by
\begin{align}
	0 &= \partial_\theta \left( \varpi \gamma v^r B^\theta \right) - \frac{4\pi}{c} \partial_\theta \left( \eta_M j_\varphi \right),	\label{radb}	 \\
	0 &= - \partial_r \left( \varpi \gamma v^r B^\theta \right) + \frac{4\pi}{c} \partial_r \left( \eta_Mj_\varphi \right),	\label{vertb}
\end{align}
where the covariant component, $j_\varphi$, of the current density reads
\begin{align}
	\frac{4\pi}{c} j_\varphi &= \frac{4\pi}{c} \left( g_{\varphi\varphi} j^\varphi + g_{\varphi t} j^t \right)	\nonumber	\\
	&= \frac{\varpi^3}{r} E^r \partial_r \omega + \frac{\varpi^2}{r^2} \partial_r \left( \alpha r B^\theta \right) - \frac{\varpi}{r} \partial_\theta B^r.
\end{align}
In the ideal MHD approximation, the magnetic diffusivity vaniushes, $\eta_M = 0$, so that
\begin{align}
	0 &= \partial_\theta \left( \varpi \gamma v^r B^\theta \right),	\\
	0 &= - \partial_r \left( \varpi \gamma v^r B^\theta \right).
\end{align}
Eqs. (\ref{radb}), (\ref{vertb}) allow for a simple solution for $B^\theta$,}
\begin{equation}
	B^\theta = \frac{\mathcal B_0}{\varpi \gamma v^r} \; , \;\; \mbox{where} \;\; \mathcal B_0 = \left. \varpi \gamma v^r B^\theta \right|_{r=\rm{r_{tr}}}.	\label{btheta}
\end{equation}
Thermal equipartition of the poloidal MF at $\rm{r_{tr}}$ serves as boundary condition, thus
\begin{equation}
	B^\theta (\rm{r_{tr}}) = \sqrt{ \frac{2 \mu_0}{\Gamma - 1} \frac k {\mu m_p} \rho_{sd}(\rm{r_{tr}}) T_{sd}(\rm{r_{tr}})},
\end{equation}
where $\Gamma$ is the adiabatic index and $\mu$ the mean molecular weight of the plasma particles. $\rho_{sd}(\rm{r_{tr}})$, $T_{sd}(\rm{r_{tr}})$ correspond to density and temperature of the standard disk at $\rm{r_{tr}}$. The radial component $B^r$ can then be derived by means of the solenoidal condition ("$\vec\nabla \cdot \vec{B} = 0$") {of the MF},
\begin{equation}
	\partial_r \left( r \varpi B^r \right) + \partial_\theta \left( \frac{r}{\alpha} B^\theta \right) = 0.	\label{maxhom}
\end{equation}
{In the BL, though, the radial component of the MF is negligible to the vertical one. Moving to higher latitudes, however, the poloidal MF will become increasingly radial due to deformation by the wind in the TZ.} The profile of the toroidal MF can be derived from {the toroidal component of the induction equations,}
\begin{multline}
	0 = \frac{\sqrt\Delta}{r^2} \partial_r \left( \alpha r \left( v^\varphi B^r - v^r B^\varphi \right) \right) +	\\
	+ \frac{\alpha}{r} \partial_\theta \left( v^\varphi B^\theta \right) + \frac{\varpi\sqrt\Delta}{r} B^r \partial_r \omega +	\\
	+ \frac{\sqrt\Delta}{r^2} \partial_r \left( \frac{\alpha \eta_M}{\gamma} \partial_r \left( \sqrt\Delta B^\varphi \right) \right) + \frac{\alpha}{r^2} \partial_\theta \left( \frac{\eta_M}{\gamma} \partial_\theta B^\varphi \right),	\label{toroidalmagneticfield}
\end{multline}
We assume that $B^\varphi$ is already very strong at $\theta = \theta_d$ so it can represent the toroidal field in the TZ while the ideal MHD approximation is still valid. Making the approximations $\partial_r \approx 1/r$ and $\partial_\theta \approx r/H_d$, one obtains from Eq. (\ref{toroidalmagneticfield})
\begin{equation}
	B^\varphi \equiv B^\varphi \Big|_{\theta = \theta_d} = \left( \frac{\omega \varpi}{\alpha v^r_d} + \frac{v^\varphi_d}{v^r_d} \right) B^r + \frac{r}{\sqrt\Delta} \frac{r}{H_d} \frac{v^\varphi_d}{v^r_d} B^\theta.	\label{bphi}
\end{equation}
The last term dominates the toroidal field at several gravitational radii whereas the first term, containing the frame-dragging potential $\omega$, becomes most dominant in the immediate vicinity of the event horizon, where the poloidal field is predominantly radial.

{In order to construct a reasonable profile for the angular velocity, we follow the results of the theoretical and numerical investigation of \citet{hujeirat2002,hujeirat2003}. Accordingly, the strong poloidal magnetic fields in the boundary layer, i.e. interior to $r_{tr}$,
extract a significant fraction of angular momentum from the plasma in the equatorial region and forcing it to rotate sub-Keplerian. The
deposited angular momentum in the TZ enables the plasma to rotate super-Keplerian, which susequently starts to accelerate its outward-oriented
motion on the dynamical time scale. In the vicinity of the event horizon, the dynamical time scale is too short to maintain thermal coupling
of the electrons with the protons. This gives rise to the formation of two-temperature proton-dominated gravitationally unbound plasma.
The profile of the angular velocity in his Newtonian analysis appears to fit well to a radial distribution of the form:
 $\Omega \propto r^{-5/4}.$ \\
In the context of general relativity however, the spin of the black hole determines the rotational behavior of the plasma in the vicinity of
the event horizon. Therefore, assuming the accretion flow to be in co-rotation with the black hole, a reasonable modification of
the above-mentioned Newtonian profile, would be:}

\begin{equation}
	\Omega = \frac{\sqrt{GM}}{r^{5/4} \rm{r_{tr}}{}^{1/4} + \rm{r_g}{}^{3/2} a}.	\label{omegabl}
\end{equation}
See \citet{diplom} for a detailed discussion of other possible profiles. This profile satisfies $\Omega(\rm{r_{tr}}) = \Omega_K(\rm{r_{tr}})$, where $\Omega_K = \sqrt{GM} \cdot ( r^{3/2} + \rm{r_g}{}^{3/2} a )^{-1}$ corresponds to the Keplerian angular velocity for co-rotating orbits. The modification to counter-rotating orbits is very straight forward.\\
For fast rotating black holes there will be a radius $\rm r _*$ where the matter in the BL is non-rotating with respect to ZAMO,
\begin{equation}
	\Omega(r_*) = \omega(r_*) \quad \Leftrightarrow	\quad	\tilde\Omega(r_*) = 0,
\end{equation}
where $\tilde\Omega = \Omega - \omega$ is the angular velocity as measured by ZAMO. Correspondingly, $\rm r _*$ is the largest, rational root of the equation
\begin{equation}
	\frac{r^2}{\rm{r_g}{}^2} + a^2 - 2 a \left( \frac{\rm{r_{tr}}{} r}{\rm{r_g}{}^2} \right)^{1/4} = 0.
\end{equation}
Our assumption is that, interior to $\rm r _*$, the matter in the BL is freely falling,
\begin{equation}
	\Omega \equiv \omega	\quad \mbox{for} \quad r \leq r_*,
\end{equation}
which implies that the inflowing matter keeps rotating with the frame-dragging frequency $\omega$, relative to the coordinate frame. On the other hand, MFs are still deformed by the frame-dragging effect, thus extracting angular momentum. The rotational energy is not extracted from the matter then, but directly from the central black hole \citep[see][for further details]{Punsly1990, Punsly2009}. Yet, the total amount extracted in this way will be negligible compared to the total rotational energy of the black hole so no spin-down is taken into account. Instead the frame-dragging potential is treated as an infinite reservoir of rotational energy.\\
{In the TZ the matter is rotating with super-Keplerian angular velocity adopting the same radial profile as in the BL. The $\Omega$-profiles in the TZ and BL are summarized,}
\begin{align}
	\Omega_d &= \left\{
\begin{array}{cl}
	\Omega_K	&	r_{tr} \leq r	\\
	\sqrt{GM} \left( r^{5/4} r_{tr}{}^{1/4} + r_g^{3/2} a \right)	&	r_* \leq r \leq r_{tr}
\end{array}
 \right.	\\
 \Omega_w &= \sqrt{GM} \left( r^{5/4} r_B{}^{1/4} + r_g^{3/2} a \right)	\quad\;\;	r_B \leq r	\label{omegatl}
\end{align}
{where the inner boundary $\rm r _B$ of the TZ is defined as the radius where centrifugal and gravitational acceleration are balanced and the effective gravity vanishes, which implies that $\Omega(r_B) = \Omega_K(r_B)$.}\\
\\
{We will now derive the $v^r$-profile from the radial momentum equation,}
\begin{multline}
	\frac{1}{2} \partial_r \left( \gamma v^r \right)^2 = \frac{1}{2} \gamma^2 c^2 \mathcal C - \frac{\partial_r P}{\rho} -	\\
	- \frac{1}{8\pi \rho } \left[ \frac{\partial_r \left( \alpha^2 r^2 {B^\theta}^2 \right)}{\alpha^2 r^2}  + \frac{ \partial_r \left( \Delta {B^\varphi}^2 \right)}{\Delta} - \frac{ \partial_r \left( r^2 \varpi^2 {E^r}^2 \right)}{r^2 \varpi^2} \right] +	\\
	+ \frac{1}{4\pi \rho } \left[ \frac{B^\theta \partial_\theta B^r}{\sqrt\Delta} + \frac{E^r \partial_\theta E^\theta}{\sqrt\Delta} - \frac{\varpi B^\theta E^r \partial_r \omega}{\alpha} \right],	\label{radialmomentumequation}
\end{multline}
{where}
\begin{equation}
	\mathcal C = - \left( 1 - \frac{{v^\varphi}^2}{c^2} \right) \partial_r \ln \alpha^2 + \partial_r \frac{{v^\varphi}^2}{c^2} - 2 \frac{{v^\varphi}^2}{c^2} \frac{\partial_r \Omega}{\tilde\Omega}.
\end{equation}
Regarding that the first term on the right-hand side is of order $c^2/r$, one may neglect the pressure term, which is of order $c_S{}^2 / r$. {Further, we will consider the magnitude of the electric field components. From Ohm's law, Eq. (\ref{ohmslaw}), we obtain}
\begin{align}
	c E^r &= v^\varphi B^\theta + \frac{\eta_M}{\gamma} \frac{1}{r} \partial_\theta B^\varphi,	\label{radialelectricfieldequation}	\\
	c E^\theta &= v^r B^\varphi - v^\varphi B^r - \frac{\eta_M}{\gamma} \frac{1}{r} \partial_r \left( \sqrt\Delta B^\varphi \right),	 \label{verticalelectricfieldequation}
\end{align}
{where $\eta_M$ is the magnetic diffusivity  (see Eq. \ref{ohm}). As the plasma motion in the TZ is said to be turbulent
 and dissipative,
 the finite width of the TZ, $\rm H_\mathrm{w},$ together with the Alf$\grave{v}$en speed, $\rm v_\mathrm{A},$ can be used to set an upper limit for the
 turbulent magnetic diffusivity as follows:  $\rm \eta_M \leq H_\mathrm{w} v_\mathrm{A},$ where $v^\mathrm{tur}\approx \rm v_\mathrm{A}/\gamma.$ Inserting $\eta_M$ in the equations, we obtain: }
\begin{align}
	c E^r &\approx v^\varphi B^\theta + v^{tur} B^\varphi,	\label{erstrength}	\\
	c E^\theta &\approx v^r B^\varphi - v^\varphi B^r + \frac{H_w}{r} \frac{\sqrt\Delta}{r} v^{tur} B^\varphi		\label{ethetastrength},
\end{align}
  Since eqs. (\ref{erstrength}) and (\ref{ethetastrength}) show that all contributions from electric fields in Eq. (\ref{radialmomentumequation}) are smaller than those from MFs, it is sufficient to show that all magnetic terms can be neglected compared to the {net} centrifugal and gravitational terms. {In this case the centrifugal and gravitational terms are the only relevant contribution to the radial acceleration so they are of order $({v^r})^2/r$. Correspondingly, this assumption is justified if $v^r$ is sufficiently larger than typical velocities associated with the electromagnetic contributions.}\\
It follows from the previous discussion that the MF is dominated by $B^\theta$ in the BL and $B^\varphi$ in the TZ. Regarding that magnetic braking operates on the dynamical time scale, i.e.
\begin{equation}
	t_{dyn} = \frac{r}{v^\varphi_d} = \frac{H_d}{v_A^\theta} = t_A,	\label{magbraktimescales}
\end{equation}
the magnitude of $B^\theta$ can be estimated. Eq. (\ref{magbraktimescales}) implies that
\begin{equation}
	v_A^\theta = \frac{H_d}{r} v^\varphi_d \ll v^\varphi_d.	\label{magnetic-fields-estimates}
\end{equation}
Following \citet{alfvenspeed}, the relativistic formula for the Alf\'en speed reads
\begin{equation}
	\gamma_A{}^2 \frac{v_A{}^2}{c^2} = \frac{B^2}{4\pi \rho \mathcal H},	\label{alfvspeed}
\end{equation}
where $\gamma_A := ( 1 - v_A{}^2/c^2 )^{-\frac{1}{2}}$. Regarding that $\mathcal H \approx c^2$ and inserting Eq. (\ref{magnetic-fields-estimates}) yields
\begin{align}
	\frac{{B^\theta}^2}{4\pi \rho_d} = \frac{\left( \frac{H_d}{r} \right)^2 {v^\varphi_d}^2}{1 - \left( \frac{H_d}{r} \right)^2 \frac{{v^\varphi_d}^2}{c^2}}	\ll {v^\varphi_d}^2.	\label{alfvenestimate}
\end{align}
Considering that the largest electromagnetic term in Eq. (\ref{radialmomentumequation}) is of the order ${v_A^\theta}^2/r$, one can see that all electromagnetic terms can be neglected compared to gravitational and centrifugal terms within the BL, {provided that $v^r >> v_A^\theta$}.\\
In the TZ the largest electromagnetic term is of the order ${v_A^\varphi}^2/r$. The stationarity condition in the TZ requires $\gamma_A v_A^\varphi = \gamma_w v^\varphi_w$ \citep{hujeirat2004}. Yet, the toroidal field is unlikely to be the main driving force since it has turning points in its vertical profile. The centrifugal and gravitational forces on the other hand remain strong throughout the TZ.
{We note that the main effect of EMF both in the outer and in the inner part of the accretion disk is to redistribute angular momentum efficiently,
so to enable a stable accretion of matter. The effect of MFs in the inner disk is to enable angular momentum exchange of two vertically
neighboring zones, rather than radially as in the case of normal accretion disk. In both cases however, MFs have the role of mediator and
therefore their corresponding total energy must remain well below the gravitational and rotational energies. }

Defining the auxiliary function $\mathcal F$ by
\begin{equation}
	\partial_r \ln \mathcal F = \frac{2 {v^\varphi}^2}{c^2 - {v^\varphi}^2} \frac{\partial_r \Omega}{\tilde\Omega},	\label{aux}
\end{equation}
one may rewrite Eq. (\ref{radialmomentumequation}) to yield
\begin{multline}
	\partial_r {v^r}^2 = {v^r}^2 \partial_r \ln \left( \alpha^2 \mathcal F \left( 1 - \frac{{v^\varphi}^2}{c^2} \right)^2 \right) -	\\
	- \left( c^2 - {v^\varphi}^2 \right) \partial_r \ln \left( \alpha^2 \mathcal F \left( 1 - \frac{{v^\varphi}^2}{c^2} \right) \right).	 \label{ordinaryvr}
\end{multline}
Eq. (\ref{ordinaryvr}) is a first order ordinary differential equation for ${v^r}^2$. Its general solution reads
\begin{equation}
	{v^r}^2 = c^2 - {v^\varphi}^2 - \frac{c^2}{\gamma_0^2} \frac{\alpha^2}{\alpha_0^2} \mathcal F \left( \frac{c^2 - {v^\varphi}^2}{c^2 - {v^\varphi_0}^2} \right)^2	\label{generalvr},
\end{equation}
where $\mathcal F$ has been normalized to $\mathcal F_0 = 1$, and where the index "0" indicates that the quantity is to be taken at radius $\rm r _0 = r_{tr/B}$ in the BL or TZ, respectively. The explicit expression of the solution to Eq. (\ref{aux}) has been moved to appendix \ref{profiles}. It is worth noting that interior to $\rm r _*$ the matter in the BL is set to be in free fall, implying $v^\varphi_d = 0$ and $\mathcal F_d = const$. The radial velocity then reduces to
\begin{equation}
	{v^r_d}^2 = c^2 - \frac{c^2}{\gamma_{d,tr}{}^2} \frac{\alpha^2}{\alpha_{tr}{}^2} \mathcal F_{d,*} \left( 1 - \frac{{v^\varphi_{d,tr}}^2}{c^2} \right)^{-2}.
\end{equation}
\begin{figure}[ht]
 \centering
\includegraphics[angle=-90,width=.5\textwidth]{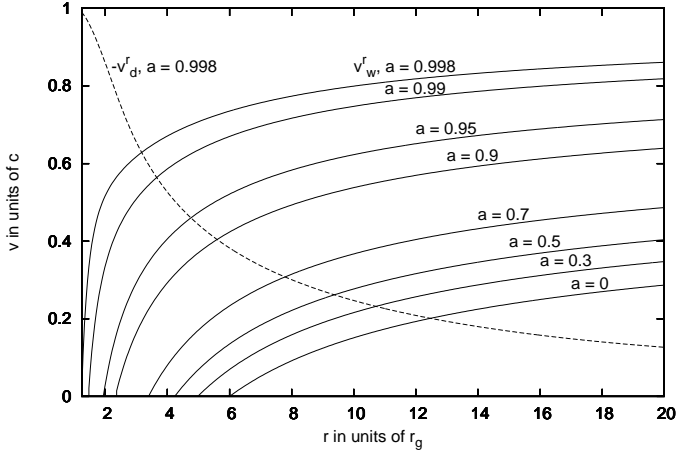}
\caption{ The radial distribution of the velocity $v^r_w$  of the outflowing plasma in the transition zone at the
distance $\rm{r_{tr}} = 40\,\rm{r_g}$ from the event horizon is plotted for different values of the spin paramter "a".
The strong dependence of the wind velocity on the spin parameter of the BH in the TZ  is most obvious in the vicinity of the event horizon,
where the baryons are strongly accelerated outward in order to overcome the deep gravitational
 well of the BH.
The dashed line corresponds to the profile of the radial velocity $v^r_d$ in the BL, which depends weakly on "a".
}
	\label{fig:vr}
\end{figure}
We proceed by deriving the profiles of $\rho_d$ from Eq. (\ref{mdotdisk}),
\begin{equation}
	\rho_d = - \frac{\dot M_d}{4\pi H_d \sqrt\Delta \gamma_d v^r_d},	\label{rhod}
\end{equation}
where $\dot M_d$ = $\dot M_d(r)$, and $\rho_w$ from the stationarity condition for the toroidal MF in the TZ, namely the equipartition of kinetic energy with the EMF, $e_{kin,w} = e_{EM,w}$, where $e_{kin/EM,w}$ is the corresponding energy density in the TZ. Inspection of the 00-component of the stress energy tensor (\ref{ergmom}) yields
\begin{align}
	e_{kin,w} &= \left(\gamma_w^2 - 1\right) \rho_w c^2,	&	e_{EM,w}	&\approx \frac{{B^\varphi}^2}{8\pi} \left( 1 + \frac{{v^r_w}^2}{c^2} \right),
\end{align}
where we have approximated the EMF in the TZ by its most dominant components $B^\varphi$ and $E^\theta$ and used Eq. (\ref{ethetastrength}) to set $c E^\theta \approx B^\varphi v^r_w$. The density in the TZ then reads
\begin{equation}
	\rho_w = \frac{{B^\varphi}^2}{8\pi c^2} \frac{1 + {v^r_w}^2/c^2}{\gamma_w^2 - 1}.	\label{rhow}
\end{equation}
In the TZ the plasma is subject to extensive centrifugal forces. We will now investigate what kind of forces are opposing this collapse and how the geometrical thickness of the TZ {can be determined from the vertical momentum equation},
\begin{multline}
	\theta \gamma^2 \left( \frac{\rm{r_g}^2 a^2 {v^r}^2}{r^2} - \frac{{v^\varphi}^2}{\mathcal S^2} \right) = \frac{1}{\rho} \partial_\theta \left[ P + \frac{{B^r}^2 + {B^\varphi}^2 - {E^\theta}^2}{8\pi} \right] -	\\
	- \frac{1}{4\pi\rho} \left[ \frac{\varpi}{r} B^r \partial_r \left( r \alpha B^\theta \right) + \frac{\alpha}{r} E^\theta \partial_r \left( r \varpi E^r \right) + \varpi^2 B^r E^r \partial_r \omega \right],	\label{vertical-momentum-equation}
\end{multline}
{where}
\begin{equation}
	\frac{1}{\mathcal S^2} = 1 + \frac{\rm{r_g} a}{r^2} \frac \omega c \left( \left( r^2 + \rm{r_g}^2 a^2 \right) \left( 1 + \frac{c^2}{{v^\varphi}^2} \right) - 2 \rm{r_g} a \sqrt\Delta \frac c {v^\varphi} \right).	\label{subs}
\end{equation}
The vertical momentum equation (\ref{vertical-momentum-equation}) reveals several forces which are due to:
\begin{enumerate}
	\item gas pressure $P_g$
	\item turbulent pressure $P_{tur}$
	\item magnetic pressure due to the toroidal MF ${B^\varphi}^2/(8\pi)$
	\item pressure and tension due to the poloidal EMF
\end{enumerate}
\begin{figure*}[ht]
\begin{minipage}{.49\textwidth}
\includegraphics[width=1.\textwidth]{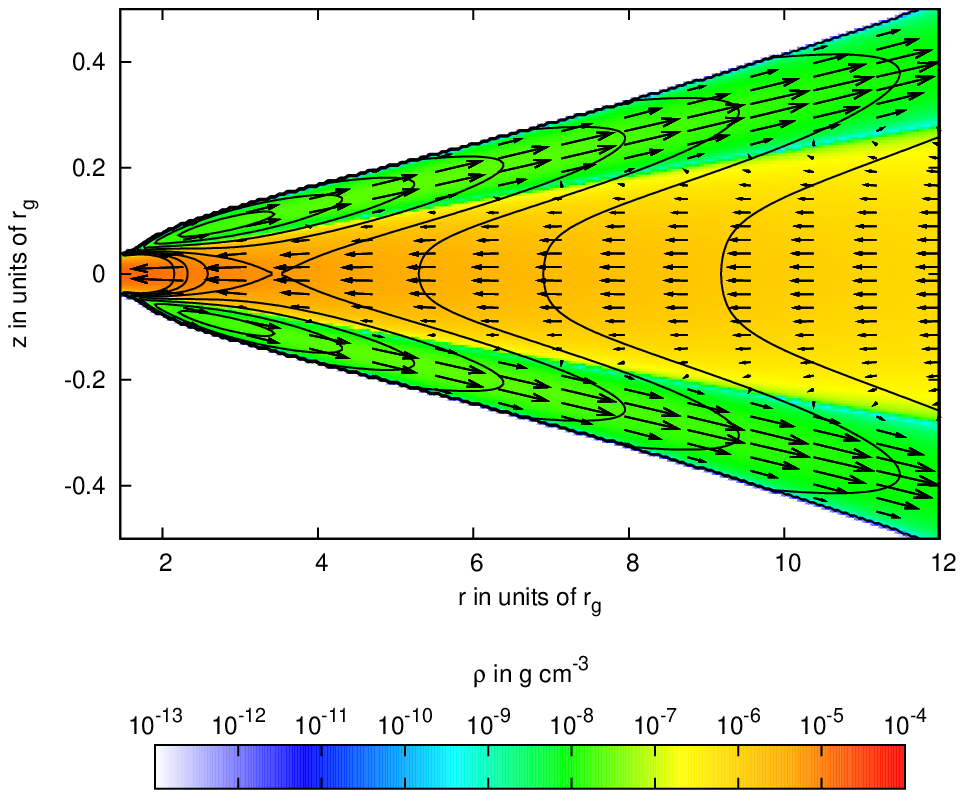}
	\caption{The two-dimensional distribution of the density, poloidal- and angular velocity in the region are displayed, where the interaction
of the inflow of matter from the accretion disk, the accreting black hole and the outflow in the TZ is most effective.
 The poloidal velocity distribution $\vec{v}_p = (v^r,v^\theta)$ in this plot  is overplotted as arrows on the color coded density distribution.
 The solid contours correspond to the angular velocity $\Omega$, which attains local maxima  in the transition
 layer, TR, where the outward-acceleration of the baryons is most significant.}
	\label{fig:rho-v-omega}
\end{minipage}	\hspace{3mm}
\begin{minipage}{.49\textwidth}
\includegraphics[width=1.\textwidth]{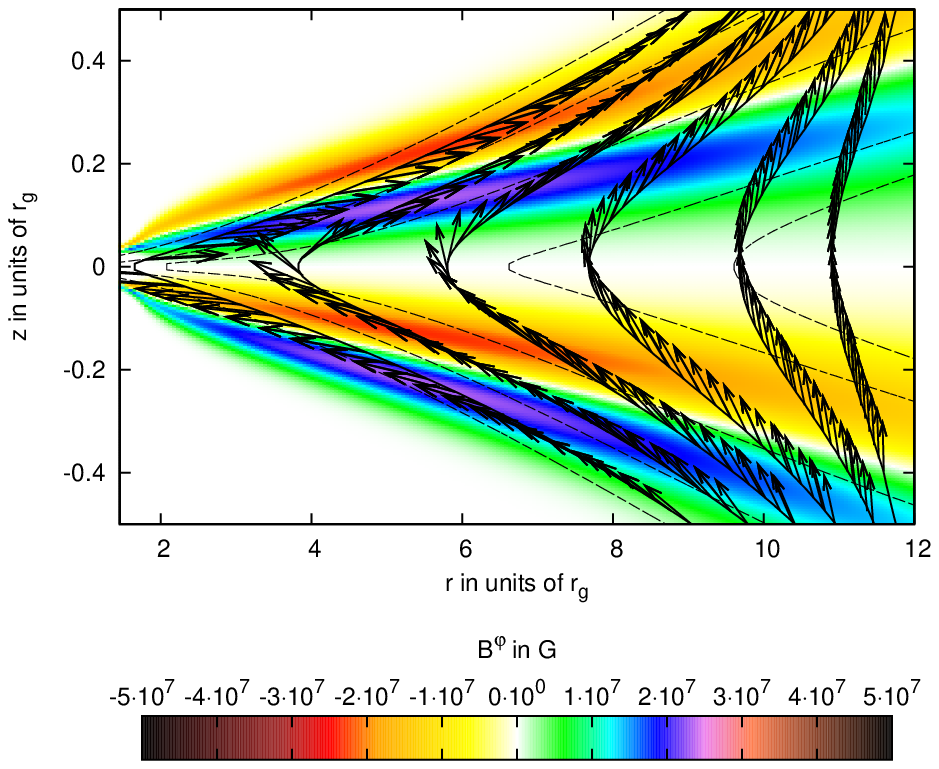}
		\caption{This plot shows the two-dimensional distribution of the MF components. The direction of the poloidal field lines is indicated by normalized arrows along solid lines of constant $A^3$, where $A = A^a e_a$ is the four-vector potential. Dashed lines mark curves of constant $|\vec B_p|$, where $\vec B_p = (B^r,B^\theta)$. {Due to the motion of the plasma the magnetic field changes its topology from dipole in the outer region into almost monopole in the vicinity of the event horizon, where the spatial variation of the toroidal field $B^\varphi$ is most significant. Correspondingly, the velocity of the wind in the TZ is nearly parallel to the field lines.}}
	\label{fig:magnetic-field-3d}
\end{minipage}
\end{figure*}

First we neglect ${v^r}^2 \rm{r_g}^2 a^2/r^2$ compared to the centrifugal term ${v^\varphi}^2/\mathcal S^2$ in Eq. (\ref{vertical-momentum-equation}), since the former is damped by a factor $\rm r ^{-2}$. Super-Keplerian motion renders gas pressure as well as all electromagnetic terms except ${B^\varphi}^2/(8\pi)$ negligible compared to centrifugal forces. However, $P_{tur}$ is most likely responsible for opposing the TZ-collapse. This is due to the fact that the toroidal MF has a turning point in the TZ giving rise to reconnection of toroidal flux tubes \citep{hujeirat2004}. Approximating $\partial_\theta$ by $-r/H_w$ and $\theta \approx H_w/r$ and writing $P_{tur}/\rho_w = \gamma_w{}^2 v_{tur}{}^2$, one obtains from Eq. (\ref{vertical-momentum-equation}) the relation
\begin{equation}
	\frac{H_w} r = \frac{v_{tur}}{v^\varphi_w} \mathcal S	\label{tlheight}.
\end{equation}
For the TZ to be stationary, we require that the amplification time scale of $B^\varphi$, $t_{amp}$, has to equal the dissipation time scale $t_{diss}$. Inspecting Eq. (\ref{toroidalmagneticfield}) yields
\begin{equation}
	t_{amp} = \frac{H_w}{\alpha v^\varphi_w} \frac{B^\varphi}{B^\theta}	= H_w{}^2 \frac{\gamma_w}{\alpha \eta_M} = t_{diss}.	 \label{tltimescales}
\end{equation}
Writing the magnetic diffusivity $\eta_M$ as
\begin{equation}
	\eta_M = H_w \gamma_w v_{tur}
\end{equation}
and using eqs. (\ref{tlheight}) and (\ref{tltimescales}) yields
\begin{equation}
	\frac{B^\theta}{B^\varphi} = \frac{H_w}{r \mathcal S}.	\label{tlheight2}
\end{equation}
Thus, the thickness of the TZ is directly related to the strength of the MF components.
Since $B^\theta/B^\varphi \ll 1$, we conclude that the TZ must be  geometrically thin.\\\\

{We note that the poloidal magnetic field remains confined to the TZ and would not diffuse throughout the corona.
This can be verified by comparing the dynamical time scale in the TZ $(\tau^{TZ}_{dyn})$ to the diffusion time
scale of the PMF into the corona
 $\tau^{PMF}_{diff}$:}
\[
\DD{\tau^{TZ}_{dyn}}{\tau^{PMF}_{diff}} \approx  \DD{({r_{tr}}/{v^{\varphi}_{TZ}})}{({r^2_{tr}}/{\eta^{mag}_{corona}})}
\approx \DD{{v}_A}{v^{\varphi}_{TZ}},
\]
{where the transition radius $r_{tr}$ has been taken as a characteristic length scale both for
the dynamical motion  and for the magnetic diffusion in the TZ, $v_A$ denotes the Alf$\grave{v}$en speed.
 As the centrifugal force acting on the super-Keplerian rotating particles in the TZ is the dominant force and therefore
 much stronger than the force due to magnetic tension that acts to straighten the magnetic force lines,
 we conclude that  ${\tau^{TZ}_{dyn}} \ll {\tau^{PMF}_{diff}}.$}\\
  Let us now derive the relative outflow rate $\dot{\mathcal M}$ {from the angular momentum equation,}
\begin{equation}
	\gamma v^r \partial_r l = \frac{1}{4\pi \rho \alpha} \left( B^r \partial_r \sqrt\Delta + B^\theta \partial_\theta \right) B^\varphi.	 \label{angularmomentum}
\end{equation}
{In order to do this, we have to integrate over the BL in the vertical direction, i.e. from $-\theta_d$ to $\theta_d$. The poloidal MF is mainly vertical in the BL and, further, at the interface between BL and TZ the vertical rate of change of the toroidal MF, $\partial_\theta B^\varphi$, exceeds the radial one, $\partial_r B^\varphi$ by a factor of order $r/H_d$ so we may savely neglect $B^r \partial_r \left( \sqrt{\Delta} B^\varphi \right)$ compared to $B^\theta \partial_\theta B^\varphi$. Then, after integration, one obtains}
\begin{equation}
	- \frac{\dot M_d}{4\pi \varpi} \frac{\partial_r l_d}{r} = \frac{B^\theta B^\varphi}{4\pi}. \label{angmom2}
\end{equation}
Using eqs. (\ref{mdotwind})-(\ref{relout}) and combining with eqs. (\ref{rhow}) and (\ref{tlheight2}) for $\rho_w$ and $H_w$ yields
\begin{align}
	\dot{\mathcal M} &= \frac{\dot M_d}{\dot M} \, g(r),	\label{explicitrelout}	\\
	g(r) &:= \frac{3}{4} \, \alpha \mathcal L \mathcal S \, \frac{\gamma_w v^r_w/c}{\gamma_w^2 - 1} \frac{1 + {v^r_w}^2/c^2}{2} \frac{l_d}{rc},	 \nonumber
\end{align}
where the function $\mathcal L$ is given in Eq. (\ref{helpl}). From this one may obtain the expressions
\begin{align}
	\dot M_d(r) &= \dot M \frac{1 - g(\rm{r_{tr}})}{1 - g(r)}, 	&	\dot{\mathcal M}(r) &= g(r) \frac{1 - g(\rm{r_{tr}})}{1 - g(r)}.
\end{align}
In order to derive the vertical drift $v^\theta$ through the surfaces of the BL one may act on the continuity Eq. (\ref{continuityequation}) with $\int_{- \theta_d}^{\theta_d} d\theta \, r^2$ to obtain
\begin{equation}
	\partial_r \left( \Sigma_d \gamma_d v^r_d \sqrt\Delta \right) = - 2 r \rho \gamma v^\theta \Big|_{\theta = \theta_d} \approx - 2 r \rho_d \gamma_d v^\theta,
\end{equation}
where $\rho$ and $\gamma$ have been approximated by their the value at the equator, $\theta = 0$, and $v^\theta = v^\theta(\theta_d)$ by its value at the interface between the disk and the TZ. Approximating $\partial_r \approx 1/r$ and using eqs. (\ref{explicitrelout}) and (\ref{mdotdisk}) yields
\begin{equation}
	v^\theta = - \frac{H_d} r \frac{\sqrt\Delta}{r} \, v^r_d \, g(r).
\end{equation}

Following \citet{hujeirat2004} the plasma in the TZ must be treated as a two-temperature flow. Due to the low density, the time scale for energy exchange is slow compared to the dynamical time scale. In the disk, on the other hand, the density is much higher implying effective energy exchange between electrons and protons. {Therefore we may use a one-temperature description in the BL of the disk while we have to consider two seperate energy equations in the TZ. The energy equation for the electrons reads}
\begin{multline}
	\rho \gamma v^r \frac{\sqrt\Delta} r \partial_r \mathcal E_e = - \frac{P_e}{r^2} \partial_r \left( r \sqrt\Delta \gamma v^r \right) +	\\
	+ \Lambda_{con} + \Lambda_{Ohm} + \Lambda_{pe} - \Lambda_{syn} - \Lambda_{B} - \Lambda_{C},	\label{ergeqe}
\end{multline}
{and the one for the protons,}
\begin{equation}
	\rho \gamma v^r \frac{\sqrt\Delta} r \partial_r \mathcal E_p = - \frac{P_p}{r^2} \partial_r \left( r \sqrt\Delta \gamma v^r \right) + \Lambda_{con} + \Lambda_{Ohm} - \Lambda_{pe},	\label{ergeqi}
\end{equation}
{where the $\Lambda_i$ are the heating and cooling functions, introduced in Sect. \ref{goveq}.} The two equations of internal energy, eqs. (\ref{ergeqe}) and (\ref{ergeqi}), are replaced by one single equation, where the contribution of ohmic heating, $\Lambda_{Ohm}$, may be dropped since the plasma is non-resistive in the BL. The main heating sources are adiabatic compression and heat conduction. The electrons cool mainly by synchrotron emission and the protons by Coulomb interaction with the electrons. We assume that heat conduction from the hot protons in the TZ suffices to compensate for the loss of heat. Then the only remaining source of heat is adiabatic compression. Therefore, in the BL, the energy equation is given by an ordinary differential equation with the simple solution
\begin{equation}
	T_d = T_d (\rm{r_{tr}}) \left( \sqrt{\frac{\Delta}{\Delta(\rm{r_{tr}})}} \frac r {\rm{r_{tr}}} \left| \frac{\gamma_d v^r_d}{\gamma_{d,tr} v^r_{d,tr}} \right| \right)^{1 - \Gamma},
\end{equation}
where $T_d$ and $\Gamma$ correspond to disk temperature and adiabatic index, respectively.
In the TZ the most dominant heating process is ohmic heating due to the large magnetic diffusivity. Electrons cool effectively by synchrotron emission. Neglecting other heating and cooling processes yields
\begin{equation}
	0 = \Lambda_{Ohm} - \Lambda_{syn},	\label{einterg}
\end{equation}
where, based on the previous discussion, $\Lambda_{Ohm}$ may be approximated by
\begin{equation}
	\Lambda_{Ohm} = \frac{4\pi}{c^2} \eta_M j^2 \approx \gamma_w v^\varphi_w \left( 1 - \frac{{v^r_w}^2}{c^2} \right) \frac{{B^\varphi}^2}{4\pi r \mathcal S}.	 \label{lambdaohm}
\end{equation}
Following \citet{relsd}, $\Lambda_{syn}$ reads
\begin{equation}
	\Lambda_{syn} = \Upsilon \frac{f_e \rho}{m_p} \left( \frac{kT}{m_e c^2} \right)^2 \frac{{B^\varphi}^2}{m_e^2}	\label{lambdasyn},
\end{equation}
where $\Upsilon = 8 \, \alpha_f{}^2 \hbar^2 / c$, $\alpha_f$ being the fine-structure constant and $f_e$ the fraction of unbound electrons to baryons.
Thus eqs. (\ref{einterg}), (\ref{lambdaohm}) and (\ref{lambdasyn}) imply that
\begin{equation}
	T_{e,w} = \frac{m_e c^2} k \sqrt{ \frac{2 m_p m_e^2} \Upsilon \frac{\gamma_w v^\varphi_w}{f_e r \mathcal S} \frac{\left( \gamma_w^2 - 1 \right) c^2}{{B^\varphi}^2} \frac{c^2 - {v^r_w}^2}{c^2 + {v^r_w}^2} }.
\end{equation}
In order to maintain stationarity, the heating and cooling processes in the TZ must operate on the same time scale as the advection of $B^\varphi$, where $t_{adv}$ can be derived from Eq. (\ref{toroidalmagneticfield}). The heating time scale is obtained by setting $t_{heat} = \rho \mathcal E/\Lambda$, where $\mathcal E_p = C_V T_{p,w}$ is the internal energy per mass of the protons, $C_V$ being the specific heat per mass. This yields
\begin{equation}
	t_{heat} = \frac{C_V \rho_w T_{p,w}}{\Lambda_{Ohm}} = \frac{r \mathcal S}{\alpha v^\varphi_w} = t_{adv}.
\end{equation}
Correspondingly, the proton temperature reads
\begin{equation}
	T_{p,w} = \frac{2 c^2}{C_V} \frac{\gamma_w}{\alpha} \left( \gamma_w^2 - 1 \right) \frac{c^2 - {v^r_w}^2}{c^2 + {v^r_w}^2}.
\end{equation}

\section{The formation and acceleration of the jet in M87}	\label{verification}

The jet in the elliptical galaxy M 87  is considered to emanate from the center, where a  $3 \cdot 10^9 M_\odot$ supermassive black
hole is believed to be residing and accreting at a rate $\dot M = 1.6 \cdot 10^{-3} \dot M_{Edd}$ from an
optically thin accretion disk that surrounds the BH \citep{jb95,dimatteo2003}. \\
The promising and most relevant feature of this jet is that the VLBI observations have restricted the size of the jet formation region in M87 to $ \rm r \leq 70\,\rm{r_g}$  from the center of the BH \citep{jb95} and that a bulk Lorentz-factor of the order
 $\gamma_{bulk} \geq 3$ is found to characterize the jet-plasma.

It should be noted, however, that the effect of the standard accretion disk in our model is expressed through the imposed
boundary conditions at the transition
radius $ \rm{r_{tr}}.$ These conditions apply also to optically thin accretions disks, as far as thermal processes are not
considered.
Therefore, without loss of generality, we may assume that the disk thickness at $ \rm{r_{tr}}$ does not differ significantly from
 $H\approx 0.1 r.$ \\
 Inside the BL, i.e. $ \rm r \leq \rm{\rm{r_{tr}}},$  magnetic braking, rather than turbulence, is the main mechanism responsible for transporting angular momentum.
 This gives rise to the formation of a transition zone between the disk and the overlying corona, where the baryons become
 gravitationally unbound due to their super-Keplerian rotation, hence start to accelerate outwards in the manner shown in
 Figs. (\ref{fig:rho-v-omega}) and (\ref{fig:magnetic-field-3d}). Additional details can be found in \citet{diplom}.

 We note that the relative outflow-rate of rest-mass $\dot{\mathcal M_{0}}$ may be estimated through the relation:
\begin{equation}
	\dot{\mathcal M_{0}} \approx \dot{\mathcal M} \, \alpha/\gamma_w
\end{equation}

\begin{figure*}[ht]
\begin{minipage}{.479\textwidth}
 \centering
	\includegraphics[width=1.\textwidth]{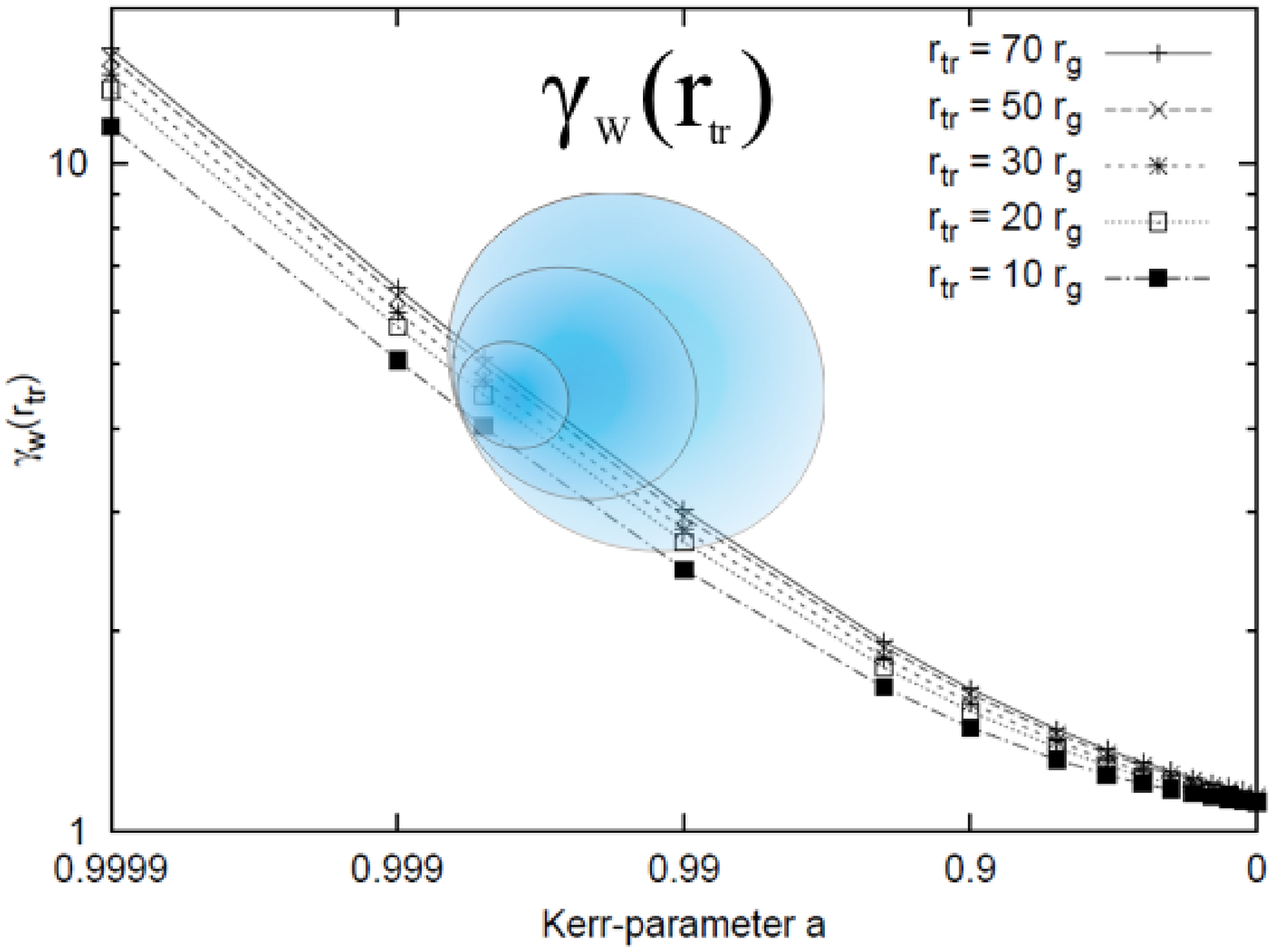}
	\caption{The Lorentz-factor $-\gamma_w$ of the outflowing plamsa in the TZ at the distance $\rm{r_{tr}}$ from the event
       horizon versus the spin of the supermassive BH  in M87 is displayed.
       The profiles show the strong correlation of $\gamma_w$ with the spin parameter of the BH, but the
       moderate correlation with the transition radius $\rm{r_{tr}}$. The correlation of the Lorentz-factor  with the spin parameter
        manifests  the argument that jets  observed with high Lorentz factor most likely emanate from the vicinity of fast rotating BHs.
       The blue region is centered with  $\gamma_w$ and  "a" values revealed by observations.
     }
	\label{fig:Gamma_M87}
	\end{minipage}	\hspace{3mm}
\begin{minipage}{.479\textwidth}
		\centering
		\includegraphics[width=1.\textwidth]{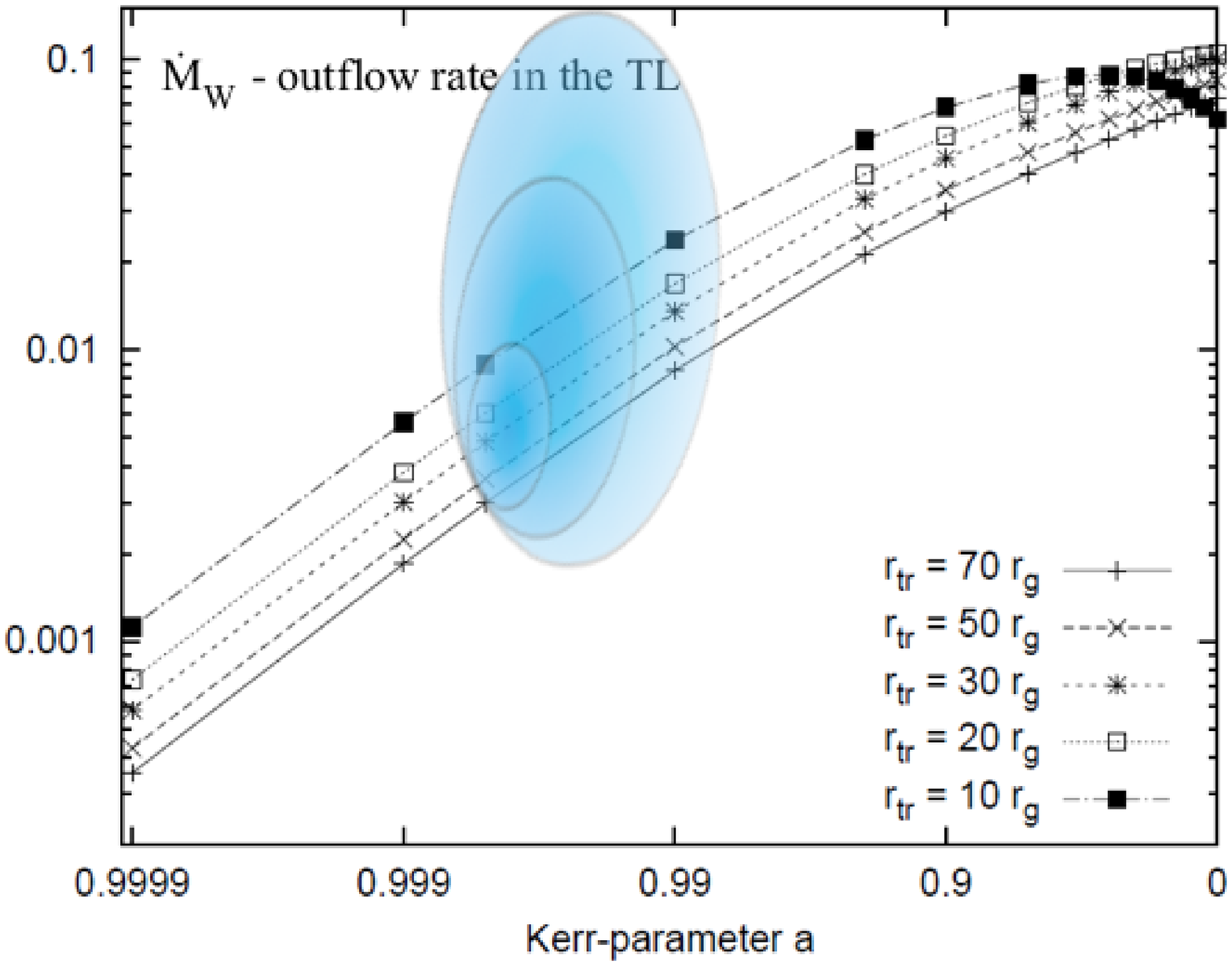}
	      \caption{ The integrated flux of the outflowing plamsa $\dot{\mathcal M_w}$ in the TZ versus
              the spin parameter of the supermassive BH  in M87 for various transition radii $\rm{r_{tr}}$ is displayed.
            The anti-correlation of $\dot{\mathcal M_w}$ with the spin parameter $a$ indicate that light
             jets most likely originate from around fast rotating BHs, whilst heavy jets from slowly rotating or Schwartzschild BHs.
           }
	\label{fig:Mdot_M87}
\end{minipage}
\end{figure*}

In Fig. (\ref{fig:Gamma_M87}, \ref{fig:Mdot_M87}) we show several profile of the Lorentz-factor $\gamma_W$ and
the outflowing rate $\dot{\mathcal M_w}$ of the plasma in the transition zone plotted as functions of the spin parameter
 $\rm a$ and for different transition radii $\rm{r_{tr}}$. The blue regions contain possible theoretical values that surround
 those revealed by from observations.\\
 Our model shows that the Lorentz factor of the gravitationally unbound baryons in the transition zone correlate
 with the spin parameter of the giant black hole. {Similarly, the TZ is best suited for explaining the
 origin and variability of the recently observed TeV and $\gamma$-ray photons from M87, which suggest that the central
 SMBH must be spinning at high rate \citep{Wang2008, Li2009}.}
  Furthermore, an accretion disk that truncate  around
 $ \rm{r_{tr}} \approx 70\,\rm{r_g}$ and a bulk Lorentz factor around 4 appear to be the most probable values that fits
 with observations. Although other truncation radii cannot be excluded, we think that $ \rm{r_{tr}} \approx 70\,\rm{r_g}$ is
 reasonable in order for the optically thin accretion disk to feed the jet with sufficient baryonic matter.\\

\section{Summary \& Conclusions}        \label{summary/outlook}

In this paper we have presented a pseudo-analytical model for the formation and acceleration of jets in the vicinity of rotating black holes under general relativistic conditions. The model is based on matching the solutions of the time-independent and axi-symmetric GRMHD equations in the three regions: the innermost disk,  the outer standard accretion disk and the transition zone between the innermost disk and the overlying corona.\\
The main aspects of our model can be summerized as follows:
\begin{enumerate}
\item In the outer region, the matter obeys the conditions of standard accretion disks.
     The magneto-rotational instability - MRI, is set to be the main driver of turbulence, which
     amplifies the magnetic fields on the dynamical time scale, $\rm{t_d}.$  $\rm{t_d}$ itself becomes shorter with decreasing
     the distance from the event horizon. A narrow transition region centered around  $\rm{r_{tr}}$  will be established,
     where magnetic and thermal energies become quantitatively comparable. \\

\item {Inside $\rm{r_{tr}},$  the collapse of the plasma in the BL causes the poloidal magnetic flux to
accumulate, suppressing therefore the generation of turbulence and diminishing heating via turbulent dissipation.
Such a development has been verified also in the 3D MHD simulations of \cite{Igumenshchev2008} and
\cite{Punsly2009}. Therefore the disk ceases to radiate as black body and turns into an observationally dark region.}  In this region, angular momentum is transported predominantly in the polar direction through magnetic braking mediated by torsional Alfven waves - TAWs.\\

\item The collapse-induced extraction of rotational energy via TAWs in the transition zone,TZ, forces the plasma to rotate super-Keplerian, hence the baryons start outward-acceleration to reach relativistic speeds already at $\rm r  \approx \rm{r_{tr}}.$
  The toroidal magnetic flux tubes ${B_{T}},$ in the TZ are magnetically unstable due to their mixed polarity. As
  a consequence, the baryons in the TZ experience an enhanced heating and acceleration through the magnetic reconnection
  of these flux tubes. The virially hot protons decouple thermally from the electrons in the TZ, due to the extensive radio
   emission by the electrons through their gyration around the magnetic field. During reconnection, both electrons and
     protons would still emit extremely hard photons whilst the bulk of kinetic energy will be carried with the protons. \\

\item The magnetically-induced collapse of the innermost disk is followed then by an inward drift of the
transition radius on the local viscous time scale. Thus, the plasma in the inner disk starts an outside-inside self-illumination through turbulent dissipation.\\

\item The procedure may repeat itself once the transition radius coincides with the event horizon of the rotating black hole.
\end{enumerate}

The parameters of the black holes, namely its mass and spin, affect both the location of the transition radius $\rm{r_{tr}},$ the
 rate of turbulence-generation in the innermost disk and the rate of acceleration of baryons in the transition zone (Fig. 3).
  The transition radius is found to decrease with increasing the spin "a" of the BH, whilst the gamma-factor of the outflowing
  baryons increases with "a". Moreover, the effective surface through which baryons enter the TZ shrinks with the spin parameter
  of the BH, implying therefore that light jets must originate from around fast spinning black holes, whilst heavy jets from
  slowly rotating or Schwarzschild black holes.

  Furthermore, the plasma in the innermost region is expected to enter the phase of turbulent-saturation on time scales that anti-correlate
  with spin of the BH as well, thus outlining the role of the spin in enhancing the MRI and therefore the generation of turbulence.\\

 When applying our model to the jet of the elliptical galaxy M87, we found that  the spin parameter
 must be near unity in order to agree with observations. Consequently, the central super-massive accreting BH in M87
 must be a maximally rotating Kerr black hole.

\vspace*{0.5cm}

\noindent{\bf Acknowledgments}  This work is supported by the Klaus-Tschira Stiftung under the
project number 00.099.2006.\\
{The anonymous referee is greatly acknowledged for carefully reading the manuscript
and for her/his valuable suggestions to improve the readability of this article.}

\clearpage

\renewcommand{\theequation}{\thesection{}\arabic{equation}}

\appendix \label{allequations}



\section{The GRMHD Equations}

The set of basis one-forms $e^a$ and vectors $e_a$, used in this work, reads
\begin{align}
	e^0 &= 	\alpha c dt,							&
	e_0 &= 	\frac{1}{\alpha c} \left( \partial_t + \omega \partial_\varphi \right),	\nonumber	\\
	e^1 &= 	\frac{r}{\sqrt\Delta} dr,	&
	e_1 &= 	\frac{\sqrt\Delta}{r} \partial_r,	\nonumber	\\
	e^2 &= 	r d\theta,								&
	e_2 &=	\frac{1}{r} \partial_\theta,	\nonumber	\\
	e^3 &= 	\varpi \left( d\varphi - \omega dt \right),	&
	e_3 &= 	\frac{1}{\varpi} \partial_\varphi.	\nonumber	\\
\end{align}
The components of the field-strength tensor in the coordinate frame may be expressed in terms of the electric and magnetic field components $E^r$, $E^\theta$, $B^r$, $B^\theta$, $B^\varphi$, measured by the ZAMO, as
\begin{align}
	F_{tr}							&=	-\frac r {\varpi} \left( E^r + \beta_{FD} B^\theta \right),	&
	F^{tr} 							&=	\frac \varpi {r} E^r,	\nonumber	\\
	F_{t\theta}				 	&=	- \alpha r \left( E^\theta - \beta_{FD} B^r \right),	&
	F^{t\theta}				 	&=	\frac	1 {\alpha r} E^\theta,	\nonumber	\\
	F_{r\theta}				 	&=	\frac{r^2}{\sqrt\Delta} B^\varphi,	&
	F^{r\theta} 				&=	\frac{\sqrt\Delta}{r^2} B^\varphi,	\nonumber	\\
	F_{\varphi r}				&=	\frac r \alpha B^\theta,	&
	F^{\varphi r} 			&=	\frac \alpha r \left( B^\theta + \beta_{FD} E^r \right),	\nonumber	\\
	F_{\theta \varphi} 	&=	r \varpi B^r,	&
	F^{\theta \varphi} 	&=	\frac 1 {r \varpi} \left( B^r - \beta_{FD} E^\theta \right),	\nonumber	\\
\end{align}
where $c \beta_{FD} = \varpi \omega/\alpha$ is the velocity of a ZAMO with respect to an observer at infinity. $F_{t\varphi}$ and hence $E^\varphi$ vanishes identically. The components of the electric current density $j^t$, $j^r$, $j^\theta$ and $j^\varphi$, as measured by ZAMO, may be expressed as
\begin{align}
	\frac{4\pi}{c} j^t &= \frac{\alpha \partial_r \left( r \varpi E^r \right)}{r^2} + \frac{\partial_\theta E^\theta}{r},	\nonumber	\\
	\frac{4\pi}{c} j^r &= \frac{\partial_\theta B^\varphi}{r},	\nonumber	\\
	\frac{4\pi}{c} j^\theta &= - \frac{\partial_r ( \sqrt\Delta B^\varphi )}{r},	\nonumber	 \\
	\frac{4\pi}{c} j^\varphi &= \frac{\varpi \partial_r \left( \alpha r B^\theta \right)}{r^2} - \frac{\partial_\theta B^r}{r} + \frac{\varpi^2 E^r \partial_r \omega}{rc}.	\label{currentdensity}
\end{align}
Next we turn to the four-velocity of the plasma, which reads
\begin{align}
	\dot x 	&= \dot t \partial_t + \dot r \partial_r + \dot \theta \partial_\theta + \dot \varphi \partial_\varphi	\nonumber	\\
	&= \gamma \left( c e_0 + v^r e_1 + v^\theta e_2 + v^\varphi e_3 \right)
\end{align}
in the coordinate- and ZAMO frame, respectively. The $v^i$ are the velocities measured by the ZAMO. On the contrary to the standard disk solution we have to include a small vertical drift $v^\theta \ll v^r, v^\varphi$. We will neglect $v^\theta$ wherever possible, though. The velocity components and Lorentz-factor are given by
\begin{align}
	v^r &= \frac r {\alpha \sqrt\Delta} \frac{dr}{dt},	&
	v^\varphi &= \frac \varpi \alpha \left( \Omega - \omega \right) \; , \quad \Omega = \frac{d\varphi}{dt},	\nonumber	\\
	v^\theta &= \frac r \alpha \frac{d\theta}{dt},	&
	\gamma &= \alpha \frac{dt}{d\tau} \; \approx \; \frac 1 {\sqrt{ 1 - {v^r}^2/c^2 - {v^\varphi}^2/c^2 }}.	\nonumber	\\	\label{velocities}
\end{align}

\section{The profiles in the BL and TZ}	\label{profiles}

We summarize the results obtained in Sect. \ref{solution}:
\begin{align}
	\Omega_d &= \left\{
\begin{array}{cl}
	\sqrt{GM} \left( r^{5/4} \rm{r_{tr}}^{1/4} + \rm{r_g}^{3/2} a \right)^{-1}	&	, \; r \in [r_*, \rm{r_{tr}}]	\\
	\omega	&	, \; r < r_*
\end{array}	\right.	\nonumber	\\
	\Omega_w &= \sqrt{GM} \left( r^{5/4} r_B^{1/4} + \rm{r_g}^{3/2} a \right)^{-1}	\nonumber	\\
	v^\varphi_d &= \frac{\varpi}{\alpha} \left( \Omega_d - \omega \right)	\nonumber	\\
	v^\varphi_w &= \frac{\varpi}{\alpha} \left( \Omega_w - \omega \right)	\nonumber	\\
	v^r_d &= - \sqrt{c^2 - {v^\varphi_d}^2 - \frac{c^2}{\gamma_{d,tr}^2} \frac{\alpha^2}{\alpha_{tr}^2} \mathcal F_d \left( \frac{c^2 - {v^\varphi_d}^2}{c^2 - {v^\varphi_{K,tr}}^2} \right)^2}	\nonumber	\\
	v^r_w &= \sqrt{c^2 - {v^\varphi_w}^2 - \frac{c^2}{\gamma_{w,B}^2} \frac{\alpha^2}{\alpha_B^2} \mathcal F_w \left( \frac{c^2 - {v^\varphi_w}^2}{c^2 - {v^\varphi_{K,B}}^2} \right)^2}	\nonumber	\\
	\gamma_d &= \gamma_{d,tr} \frac{\alpha_{tr}}{\alpha} \frac{1}{\sqrt{\mathcal F_d}} \frac{c^2 - {v^\varphi_{K,tr}}^2}{c^2 - {v^\varphi_d}^2}	 \nonumber	\\
	\gamma_w &= \gamma_{w,B} \frac{\alpha_{B}}{\alpha} \frac{1}{\sqrt{\mathcal F_w}} \frac{c^2 - {v^\varphi_{K,B}}^2}{c^2 - {v^\varphi_w}^2}	 \nonumber	\\
	\rho_d &= - \frac{\dot M_d}{4\pi H_d \sqrt\Delta \gamma_d v^r_d}	\nonumber	\\
	\rho_w &= \frac{{B^\varphi}^2}{8\pi c^2} \frac{1 + {v^r_w}^2/c^2}{\gamma_w^2 - 1}	\nonumber	\\
	B^\theta &= \frac{\mathcal B_0}{\varpi \gamma_d v^r_d}	\nonumber	\\
	B^r &= \frac{1}{r\varpi} \left( \frac{r^2}{\alpha} B^\theta - \frac{\rm{r_{tr}}{}^2}{\alpha_{tr}} B^\theta_{tr} \right)	\nonumber	\\
	B^\varphi &= \left( \frac{\omega \varpi}{\alpha v^r_d} + \frac{v^\varphi_d}{v^r_d} \right) B^r + \frac{r}{\sqrt\Delta} \frac{r}{H_d} \frac{v^\varphi_d}{v^r_d} B^\theta	\nonumber	\\
	H_w &= \frac{B^\theta}{B^\varphi} r \mathcal S	\nonumber	\\
\eta_M &= \frac{H_w^2}{r \mathcal S} \gamma_w v^\varphi_w	\nonumber	\\
\dot{\mathcal M} &= g \, \frac{1 - g_{tr}}{1 - g}	\label{outflow}	\nonumber	\\
v^\theta &= - \frac{H_d} r \frac{\sqrt\Delta}{r} \, v^r_d \, g	\nonumber	\\
T_d &= T_d (\rm{r_{tr}}) \left( \sqrt{\frac{\Delta}{\Delta(\rm{r_{tr}})}} \frac r {\rm{r_{tr}}} \left| \frac{\gamma_d v^r_d}{\gamma_{d,tr} v^r_{d,tr}} \right| \right)^{1 - \Gamma}	\nonumber	\\
	T_{e,w} &= \frac{m_e c^2} k \sqrt{ \frac{2 m_p m_e^2} \Upsilon \frac{\gamma_w v^\varphi_w}{f_e r \mathcal S} \frac{\left( \gamma_w^2 - 1 \right) c^2}{{B^\varphi}^2} \frac{c^2 - {v^r_w}^2}{c^2 + {v^r_w}^2} }.	\nonumber	\\
	T_{p,w} &= \frac{2 c^2}{C_V} \frac{\gamma_w}{\alpha} \left( \gamma_w^2 - 1 \right) \frac{c^2 - {v^r_w}^2}{c^2 + {v^r_w}^2}.	\nonumber	\\
\end{align}
\noindent The above expressions depend on the six parameters $\dot M_d$, $M$, $a$, $\rm r $, $\rm{r_{tr}}$, $\rm r _B$. We will reformulate them in terms of the non-dimensional, scaled variables
\begin{equation}
	x = \frac{r}{\rm{r_g}}	\; , \quad	m = \frac{M}{M_\odot}	\; , \quad	\dot m = \frac{\dot M_d}{10^{17}\,g\,s^{-1}},
\end{equation}
and the following constants and auxiliary functions:
\begin{align}
	\mathcal B_6 &= \frac{|\mathcal B_0|}{10^6 \, G \cdot c r_{g,\odot}} \quad (\approx 1-10 \; \mbox{ for stellar systems})	\nonumber	\\
	\mathcal B^r &= 1 - \left( \frac{x}{x_{tr}} \right)^{-1} \cdot \left( \frac{\mathcal D}{\mathcal D_{tr}} \right)^{1/2} \cdot \frac{\mathcal Q_d}{\mathcal Q_{d,tr}}	\nonumber	\\
		\mathcal B^\varphi &= \vartheta + 10^{-2} \cdot \mathcal B^r \, \mathcal H_2 \left( \vartheta + 2\,a\,x^{-7/4} \,  \, \mathcal O_d{} \, \mathcal T_d{}^{-1} \right)	\nonumber	\\
	\mathcal D &= 1 - \frac{2}{x} + \frac{a^2}{x^2}	\nonumber	\\
		\mathcal G_{d/w} &= \mathcal F_{d/w} \, \gamma_{tr/B}{}^{-2} \alpha_{tr/B}{}^{-2} \left( 1 - \frac{{v^\varphi_{K,tr/B}}^2}{c^2} \right)^{-2}	 \nonumber	\\
	\mathcal H_2 &= 10^2 \cdot \frac{H_{sd,tr}}{\rm{r_{tr}}}	\nonumber	\\
	\mathcal O_{d/w} &= x_{tr/B}{}^{1/4} + a\,x^{-5/4}	\nonumber	\\
	\mathcal Q_{d/w} &= \gamma_{d/w} \frac{|v^r_{d/w}|}{c} = \left( \mathcal G_{d/w}{}^{-1} \, \mathcal D^{-1} \, \mathcal W \, 	\mathcal V_{d/w}{}^{-1} - 1 \right)^{1/2}	\nonumber	\\
	\mathcal T_0 &= \frac{T_{sd}(\rm{r_{tr}})}{10^7 K}	\nonumber	\\
	\mathcal T_{d/w} &= 1 - 2\,a\,x_{tr/B}{}^{1/4} \, x^{-7/4} + \frac{a^2}{x^2}	\nonumber	\\
	\mathcal V_{d/w} &= 1 - \frac{{v^\varphi_{d/w}}^2}{c^2}	\nonumber	\\
	\mathcal W &= 1 + \frac{a^2}{x^2} + \frac{2 a^2}{x^3}	\nonumber	\\
	\vartheta &= \Theta(r - r_*) = \left\{
\begin{array}{ccl}
	1 & , & r \in [r_*, \rm{r_{tr}}]	\\
	0 & , & r < r_*
\end{array} \right.	\nonumber	\\
\mathcal L &= \frac{4}{3} \frac{1}{1 - \frac{{v^\varphi_d}^2}{c^2}} \cdot \Bigg[ 4 \frac{r^2 - \frac{\rm{r_g}^3}{r}
a^2}{\varpi^2} - 2 \frac{r^2 - r \rm{r_g}}{\Delta} +	\nonumber	\\
& + \tilde\Omega_d^{-1} \left( \omega \frac{3 r^2 + \rm{r_g}^2 a^2}{\varpi^2} \left( 1 + \frac{{v^\varphi_d}^2}{c^2}
\right) - \frac 5 4 \rm{r_{tr}}{}^{\frac 1 4} r^{\frac{5}{4}} \frac{\Omega_d{}^2}{\sqrt{\rm{r_g}} c} \right) \Bigg]
\nonumber	\label{helpl}	\\
	\mathcal S &= \Bigg[ 1 + \frac{2 a^2}{x^3} \mathcal W^{-1} \Bigg( - \frac{2 a}{x} \mathcal D^{1/2} \left( 1 - \mathcal
V_w \right)^{-1/2} +	\nonumber	\\
& +	 \left( 1 + \frac{a^2}{x^2} \right) \left( 1 + \left( 1 - \mathcal V_w \right)^{-1} \right) \Bigg) \Bigg]^{-1/2}
\nonumber	\\
	\mathcal F_{d/w} &= \exp\Bigg[ - \frac{10}{y_0{}^2} \sum\limits_{i=1}^{10} \prod\limits_{\stackrel{j=1}{i \neq
j}}^{10} \frac{1}{y_i - y_j} \cdot \Bigg( \sum\limits_{k=1}^{7} \frac{y^k - y_0{}^k}{k} y_i{}^{7 - k} +	\nonumber	\\
& +	\left( y_i{}^7 - 2 a y_0 + \frac{a^2}{y_0} \right) \ln\frac{y - y_i}{y_0 - y_i} - \frac{a^2}{y_i} \ln\frac{y}{y_0}
\Bigg) \Bigg], \nonumber	\\
\end{align}

where $y = (r/\rm{r_g})^{1/4}$ and $y_1, y_2, y_3$ are the three real roots of the equation
\clearpage

\begin{equation}
	y^5 - y_0{}^{-2} y^3 - 2 y + 2 a y_0{}^{-1} = 0	\label{poly},
\end{equation}

which are located in the interval $[-2,1.5]$ for all reasonable parameters. Further we have

\begin{align}
	y_{4/5} &= - \frac{y_1 + y_2 + y_3}{2} \pm	\nonumber	\\
& \pm \sqrt{y_0{}^{-2} - \frac{y_1{}^2 + y_2{}^2 + y_3{}^2}{2} - \frac{(y_1 + y_2 + y_3)^2}{4}},	 \nonumber	\\
	y_n &= - \left( \frac{a}{y_0} \right)^{1/5} e^{2\pi i n/5},
\end{align}
where $n \in \{6,7,8,9,10\}$. In terms of these variables, the solution can be further reformulated as follows
\begin{align}
\begin{array}{ll}
\Omega^{(1)}_d &=2.0 \cdot 10^5 s^{-1} \cdot m^{-1} \, x^{-5/4} \, \mathcal O_d{}^{-1}	\;~~~~~ for~~~ r \in [r_*, \rm{r_{tr}}]	\\
\Omega^{(2)}_d &= 4.1 \cdot 10^5 s^{-1} \cdot m^{-1} \, a \, x^{-3} \, \mathcal W^{-1} \;\;~~~~~ for~~~   r < r_* \nonumber\\
\Omega_w &= 2.0 \cdot 10^5 s^{-1} \cdot m^{-1} \, x^{-5/4} \, \mathcal O_w{}^{-1}	\nonumber \\
	\beta^\varphi_d &= v^\varphi_w/c = \vartheta \cdot x^{-1/4} \, \mathcal D^{-1/2} \, \mathcal O_d{}^{-1} \, \mathcal T_d	\nonumber	\\
	\beta^\varphi_w &= v^\varphi_w/c = x^{-1/4} \, \mathcal D^{-1/2} \, \mathcal O_w{}^{-1} \, \mathcal T_w	\nonumber	\\
	\beta^r_d &= -v^r_d/c = \left( \mathcal V_d - \mathcal D \, \mathcal G_d \, \mathcal V_d{}^2 \, \mathcal W^{-1} \right)^{1/2}	\nonumber	 \\
	\beta^r_w &= v^r_w/c = \left( \mathcal V_w - \mathcal D \, \mathcal G_w \, \mathcal V_w{}^2 \, \mathcal W^{-1} \right)^{1/2}	\nonumber	 \\
	\gamma_d &= \mathcal D^{-1/2} \, \mathcal G_d{}^{-1/2} \, \mathcal V_d{}^{-1} \, \mathcal W^{1/2}	\nonumber	\\
	\gamma_w &= \mathcal D^{-1/2} \, \mathcal G_w{}^{-1/2} \, \mathcal V_w{}^{-1} \, \mathcal W^{1/2}	\nonumber	\\
	\rho_d &= 1.2 \cdot 10^{-3} \frac{g}{cm^3} \cdot \dot m \, m^{-2} \, x^{-2} \, \mathcal D^{-1/2} \, \mathcal H_2{}^{-1} \, \mathcal Q_d{}^{-1}	 \\
	\rho_w &= 4.4 \cdot 10^{-7} \frac{g}{cm^3} \cdot m^{-2} \, x^{-5/2} \, \mathcal B_6{}^2 \, {\mathcal B^\varphi}^2 \,
\mathcal D^{-1} \mathcal G_w \, \mathcal H_2{}^{-2} \cdot \, 	\nonumber	\\
	&\quad\cdot \mathcal O_d{}^{-2} \, \mathcal Q_d{}^{-2} \, \mathcal T_d{}^2 \, \mathcal V_w{}^2 \, \mathcal W^{-2} \,
{\beta^r_d}^{-2}	 \left( 1 + {\beta^r_w}^{2} \right) \cdot	\nonumber	\\
	&\quad\cdot \left( 1 - \mathcal D \, \mathcal G_w
\mathcal V_w{}^2 \mathcal W^{-1} \right)^{-1}	 \nonumber	\\
	B^r &= 10^6 \, G \cdot m^{-1} \, x^{-1} \, \mathcal B_6 \, \mathcal B^r \, \mathcal D^{-1/2} \, \mathcal Q_d^{-1} \, \mathcal W^{-1/2}	 \nonumber	 \\
	B^\theta &= 10^6 \, G \cdot m^{-1} \, x^{-1} \, \mathcal B_6 \, \mathcal Q_d{}^{-1} \, \mathcal W^{-1/2}	\nonumber	\\
	B^\varphi &= - 10^8 \, G \cdot m^{-1} \, x^{-5/4} \, \mathcal B_6 \, \mathcal B^\varphi \, \mathcal D^{-1} \, \mathcal
H_2^{-1} \cdot\,	\nonumber	\\
	&\quad\cdot \mathcal O_d{}^{-1} \, \mathcal Q_d{}^{-1} \, \mathcal T_d \, \mathcal W^{-1/2} \,
{\beta^r_d}^{-1} \nonumber	\\
	H_w &= 10^{-2} \, \rm{r_g} \cdot x^{5/4} \, {\mathcal B^\varphi}^{-1} \, \mathcal D \, \mathcal H_2 \, \mathcal O_d \, \mathcal S \, \mathcal T_d{}^{-1} \, \beta^r_d	\nonumber	\\
	\eta_M &= 4.4 \cdot 10^{11} \frac{cm^2}{s} \cdot m \, x^{5/4} \, {\mathcal B^\varphi}^{-2} \, \mathcal D \, \mathcal
G_w{}^{-1/2} \, \mathcal H_2{}^2 \cdot	\nonumber	\\
&\quad\cdot \mathcal O_d{}^2 \, \mathcal O_w{}^{-1} \, \mathcal S \, \mathcal T_d{}^{-2} \, \mathcal T_w \, \mathcal
V_w{}^{-1} \, \mathcal W^{1/2} \, {\beta^r_d}^2	\nonumber	\nonumber	\\
	\dot{\mathcal M} &= 3.6 \cdot 10^{-4} \cdot \dot m^{-1} \, x^{-1/4} \, \mathcal B_6{}^2 \, \mathcal B^\varphi \,
\mathcal D^{1/2} \, \mathcal G_w \, \mathcal H_2{}^{-1} \cdot	\nonumber	\\
&\quad\cdot \mathcal O_d{}^{-1} \, \mathcal Q_d{}^{-2} \, \mathcal Q_w \, \mathcal S \, \mathcal T_d \, \mathcal V_w{}^2 \,
\mathcal W^{-2} \, {\beta^r_d}^{-1} \left( 1 + {\beta^r_w}^2 \right) \cdot	\nonumber	\\
	&\quad\cdot \left( 1 - \mathcal  D \mathcal G_w \mathcal V_w{}^2 \mathcal W^{-1} \right)^{-1}	 \nonumber	\\
	\beta^\theta &= v^\theta/c = 7.3 \cdot 10^{-6} \cdot \dot m^{-1} \, x^{1/2} \, \mathcal B_6{}^2 \, \mathcal D^3 \,
\mathcal G_w \, \mathcal O_w{}^2 \cdot	\nonumber	\\
	&\quad\cdot \mathcal Q_d{}^{-2} \, \mathcal Q_w \, \mathcal S^2 \, \mathcal T_w{}^{-2} \, \mathcal V_w{}^2 \, \mathcal
W^{-2} \, \beta^r_d	 \nonumber	 \\
	T_d &= 10^7 K \cdot \mathcal T_0 \cdot \left( \left( \frac{x}{x_{tr}} \right)^2 \cdot \left( \frac{\mathcal D}{\mathcal D_{tr}} \right)^{1/2} \cdot \frac{\mathcal Q_d}{\mathcal Q_{d,tr}} \right)^{1 - \Gamma}	\nonumber	\\
	T_{e,w} &= 7.9 \cdot 10^9 K \cdot m \, x^{5/8} \, \mathcal B_6^{-1} \, {\mathcal B^\varphi}^{-1} \, \mathcal
G_w{}^{-3/4} \, \mathcal H_2 \cdot	\nonumber	\\
	&\quad\cdot \mathcal O_d \, \mathcal O_w{}^{-1/2} \, \mathcal Q_d \, \mathcal S^{-1/2} \, \mathcal T_d{}^{-1} \,
\mathcal T_w{}^{1/2} \, \mathcal V_w{}^{-3/2} \, \mathcal W^{5/4} \cdot	\nonumber	\\
	& \cdot \beta^r_d \left( 1 - \mathcal D \mathcal G_w \mathcal V_w{}^2 \mathcal W^{-1} \right)^{1/2} \left( 1 -
{\beta^r_w}^2 \right)^{1/2} \left( 1 + {\beta^r_w}^2 \right)^{-1/2}	\nonumber  \\
	T_{p,w} &= 7.3 \cdot 10^{12} K \cdot \mathcal D^{-2} \, G_w{}^{-3/2} \, \mathcal V_w{}^{-3} \, \mathcal W^2
\cdot	\nonumber	\\
	&\quad\cdot \left( 1 - \mathcal D \mathcal G_w \mathcal V_w{}^2 \mathcal W^{-1} \right) \left( 1 - {\beta^r_w}^2
\right) \left( 1 + {\beta^r_w}^2 \right)^{-1}.	\nonumber
 \end{array}
\end{align}


\begin{thebibliography}{}

\bibitem[{{Balbus} \& {Hawley}(1991)}]{mri1}
{Balbus}, S.~A. \& {Hawley}, J.~F. 1991, \apj, 376, 214

\bibitem[{{Belloni} {et~al.}(2000)}]{Belloni2000}
{Belloni}, T.; {Klein-Wolt}, M.; {Méndez}, M.; et al.

\bibitem[{{Blandford} \& {Znajek}(1977)}]{BZ77}
{Blandford}, R.~D. \& {Znajek}, R.~L. 1977, MNRAS, 179, 433

\bibitem[{{Brezinski}(2010)}]{diplom}
{Brezinski}, F. 2010, {Diploma thesis: "A General Relativistic Model for the Formation of Jets
  in Microquasars and AGN"}, Faculty of Physics and Astronomy, University of Heidelberg, Germany

\bibitem[{{Camenzind}(2007)}]{camenzind}
{Camenzind}, M. 2007, {Compact objects in astrophysics: white dwarfs, neutron
  stars, and black holes}, ed. {Camenzind, M.}

\bibitem[{{Di Matteo} {et~al.}(2003){Di Matteo}, {Allen}, {Fabian}, {Wilson},
  \& {Young}}]{dimatteo2003}
{Di Matteo}, T., {Allen}, S.~W., {Fabian}, A.~C., {Wilson}, A.~S., \& {Young},
  A.~J. 2003, \apj, 582, 133

\bibitem[{{Fender} {et~al.} (2007)}]{FenderEtAl2007}
Fender, R.,  Koerding, E., Belloni, T., et al., 2007, astro-ph=0706.3838

	
\bibitem[{{Fragile}(2008)}]{Fragile2008}
{Fragile}, P.~C. 2008, Proceedings of Science, in "Microquasars and Beyond", Izmir, Turkey


\bibitem[{Gedalin(1993)}]{alfvenspeed}
Gedalin, M. 1993, Phys. Rev. E, 47, 4354

\bibitem[{{Hujeirat} \& {Camenzind} (2000){Hujeirat}, \& {Camenzind} }]{Hujeirat2000}
{Hujeirat}, A., {Camenzind}, M., M. 2000, A\&A, 362, L41


\bibitem[{{Hujeirat} {et~al.}(2002){Hujeirat}, {Camenzind}, \&
  {Livio}}]{hujeirat2002}
{Hujeirat}, A., {Camenzind}, M., \& {Livio}, M. 2002, A\&A, 394, L9


\bibitem[{{Hujeirat}(2003)}]{hujeirat2003-2}
{Hujeirat}, A. 2003, ArXiv Astrophysics e-prints

\bibitem[{{Hujeirat} {et~al.}(2003){Hujeirat}, {Livio}, {Camenzind}, \&
  {Burkert}}]{hujeirat2003}
{Hujeirat}, A., {Livio}, M., {Camenzind}, M., \& {Burkert}, A. 2003, A\&A, 408

\bibitem[{{Hujeirat}(2004)}]{hujeirat2004}
{Hujeirat}, A. 2004, A\&A, 416, 423

\bibitem[{{Igumenshchev}(2008)}]{Igumenshchev2008}
{Igumenshchev}, A. 2008, ApJ, 677, 317

\bibitem[{{Junor} \& {Biretta}(1995)}]{jb95}
{Junor}, W. \& {Biretta}, J.~A. 1995, \aj, 109, 500

\bibitem[{{Li} {et al. }(2009)}]{Li2009}
{Li}, Y-R.,  {Yuan}, Y-F., {Wang}, J-M., {Wang}, J-C.,  \& {Zhang}, S., 2009, \apj, 699, 513


\bibitem[{{Livio}(2009)}]{Livio2009}
{Livio}, M. 2009, {ASS-Proceedings: Protostellar Jets in Context}, ed. {Livio, M.}

\bibitem[{{Meier}(2004)}]{grohm}
{Meier}, D.~L. 2004, \apj, 605, 340

\bibitem[{{McKinney} \& {Gammie}(2004)}]{McKinney2004}
{McKinney}, J.~C., {Gammie}, C.~F., 2004, \apj, 605, 340

\bibitem[{{Novikov} \& {Thorne}(1973)}]{relsd}
{Novikov}, I.~D. \& {Thorne}, K.~S. 1973, in Black Holes (Les Astres Occlus),
  343--450

\bibitem[{{Page} \& {Thorne}(1974)}]{relsd2}
{Page}, D.~N. \& {Thorne}, K.~S. 1974, \apj, 191, 499

\bibitem[{{Punsly} \& {Coroniti }(1990)}]{Punsly1990}
{Punsly}, B. \&  {Coroniti}, F.V., 1990, \apj, 354, 583


\bibitem[{{Punsly} {et al. }(2009)}]{Punsly2009}
{Punsly}, B.,  {Igumenshchev}, I.V., \& {Hirose}, S., 2009, \apj, 704, 1065

\bibitem[{{Rothstein} \& {Lovelace}(1974)}]{Lovelace2008}
{Rothstein}, D.M. \& {Lovelace}, R.V.E.,  2008, \apj, 677, 1221


\bibitem[{{Shakura} \& {Sunyaev}(1973)}]{ssd}
{Shakura}, N.~I. \& {Sunyaev}, R.~A. 1973, A\&A, 24, 337


\bibitem[{{Wang} {et al. }(2008)}]{Wang2008}
{Wang}, B.,  {Li}, I.V., \& {Wang}, S., 2008, \apj, 676, L109



\end{thebibliography}
\end{document}